# Science as a Public Good: Public Use and Funding of Science

Yian Yin[1,2,3], Yuxiao Dong[4], Kuansan Wang[4], Dashun Wang[1,2,3,5*], Benjamin F. Jones[1,2,5,6*]

[1]*Center for Science of Science and Innovation, Northwestern University, Evanston IL*
[2]*Northwestern Institute on Complex Systems, Northwestern University, Evanston IL*
[3]*McCormick School of Engineering, Northwestern University, Evanston IL*
[4]*Microsoft Research, Redmond WA*
[5]*Kellogg School of Management, Northwestern University, Evanston IL*
[6]*National Bureau of Economic Research, Cambridge MA*
[*]*Correspondence to* dashun.wang@northwestern.edu, bjones@kellogg.northwestern.edu

**Knowledge of how science is consumed in public domains is essential for a deeper understanding of the role of science in human society. While science is heavily supported by public funding, common depictions suggest that scientific research remains an isolated or 'ivory tower' activity[1-5], with weak connectivity to public use, little relationship between the quality of research and its public use[6-8], and little correspondence between the funding of science and its public use[9-12]. This paper introduces a measurement framework to examine public good features of science, allowing us to study public uses of science, the public funding of science, and how use and funding relate. Specifically, we integrate five large-scale datasets that link scientific publications from all scientific fields to their upstream funding support and downstream public uses across three public domains -- government documents, the news media, and marketplace invention. We find that the public uses of science are extremely diverse, with different public domains drawing distinctively across scientific fields. Yet amidst these differences, we find key forms of alignment in the interface between science and society. First, despite concerns that the public does not engage high-quality science, we find universal alignment, in each scientific field and public domain, between what the public consumes and what is highly impactful within science. Second, despite myriad factors underpinning the public funding of science, the resulting allocation across fields presents a striking alignment with the field's collective public use. Overall, public uses of science present a rich landscape of specialized consumption, yet collectively science and society interface with remarkable, quantifiable alignment between scientific use, public use, and funding.**

Science is often seen to provide substantial impacts beyond the community of scientists themselves — for technological progress, government function, basic human curiosity, and more[13-19]. Given the potential benefits, many nations have built institutional architectures to support science through public investment, following the logic of public goods[20-22]. Like a public park, which is funded by the government and can be visited for free, scientific research is substantially funded by governments with its results placed in the public domain. This institutional design, which depends upon taxing the public (rather than market-based mechanisms), relies on the idea that public investment in science can match the public interest in science.

Although the idea of science as a public good is foundational to the scientific ecosystem[21-23], empirically testing the key premise of alignment between public funding and public use has remained elusive, mainly due to the difficulty in collecting systematic data. Moreover, the lack of measurement has invited substantial skepticism. Indeed, many observers view scientific research as a cloistered or 'ivory tower' activity that rarely corresponds to the public interest[1-5]. For example, the "two communities" and "two cultures" theories highlight substantial knowledge and interest gaps between scientists and policymakers, disconnecting scientific research from policy insights[7,24-26] and suggesting little relationship between the quality of research and its public usage[6-8]. Meanwhile, scientists may have peculiar interests, with little exposure to real world problems or incentives to tackle them[18,27]. These potential gaps further animate root concerns over the public funding of science and its proper allocation[9-12]. For example, policymakers have long criticized the National Science Foundation (NSF) for funding frivolous research and have called for greater transparency around the relevance of science[9,10]. Some prominent academics and commentators, including Nobel-Prize winner Milton Friedman, have taken the position that the government should not fund science, favoring purely private sector research instead[11,12].

This paper introduces a measurement framework to examine public good features of science, allowing us to study public uses of science, the public funding of science, and how public use and public funding relate. Building on prior research that considers the use of science within a given public domain[28-33], here we integrate five large-scale datasets that link scientific publications from all scientific fields to their upstream funding support and downstream public uses across three public domains. Our first dataset ($D_1$) is scientific publications, using the Microsoft Academic Graph (MAG)[34], which is one of the largest bibliometric databases of scientific research in the world (SI S1.1). Our second dataset ($D_2$) leverages the Microsoft Bing search engine to collect over 6 million government documents available online across all branches of the U.S. government[35]. Using a machine reading technology, we systematically identify academic publications that are referenced in these government documents and match these references to the MAG. This novel pipeline allows us to collect a wide-scale dataset on how government documents consume scientific knowledge (SI S1.2). In total, we identify 353,977 unique academic publications cited by 39,618 government documents. Our third dataset ($D_3$) uses the Altmetric data[29,30] to track academic publications covered by mainstream media reports. Matching these publications to the MAG data yields 548,962 unique papers covered by 2,417 media outlets (SI S1.3). Building on prior work[31-33], our fourth dataset ($D_4$) links all patents granted by the United States Patent and Trademark Office (USPTO) to the academic papers they reference, yielding 2,726,562 papers cited by 1,314,474 patents. Here we focus on papers published between 2005 and 2014, a common period covered by all three datasets, resulting in 116,462, 216,657, and 764,553 papers cited in government, news, and patent documents, respectively. Finally, we integrate funding records, using a novel dataset ($D_5$), Dimensions[36], which includes 5 million projects funded by 400 funding agencies worldwide and links each funded project with its resulting publications. The SI further details the construction of each dataset and additional validations (SI S1-S2).

Our first analyses measure the usage of scientific research in the three public domains. To conduct this analysis, we first leverage the MAG's classification of papers across 19 top-level fields. To account for cross-field differences in publication volume, we define a Relative Consumption Index, *RCI*. For a given public domain (*d*) and field (*f*), *RCI* measures the fraction of papers in the field consumed by that public domain, normalized by the same fraction calculated on all fields for that domain. That is,

$$RCI_d^f = \frac{\text{\# papers in field } f \text{ consumed by domain } d / \text{ \# papers in } f}{\text{Total \# papers consumed by domain } d / \text{ Total \# papers}}.$$

We find that the public uses of science are diverse, with many fields showing substantially specialized usage in public domains (Fig. 1). First consider science and engineering. Computer science, materials science, and mathematics (Fig. 1d, i-j) present substantially larger *RCI* values for patents than for government or news. By contrast, environmental science and geology (Fig. 1f, h) contribute relatively strongly in government and media documents compared to patents. Finally, physics, chemistry, medicine, and biology present a broader range of use (Fig. 1b-c, k-l). Among all fields, biology is the only one over-represented across all three channels, demonstrating a uniquely general relevance to these broad domains beyond science.

Social sciences, by contrast, exhibit a visibly different pattern of public use. The social sciences are strongly consumed in government and media domains while showing systematically low usage in patents (Fig. 1m-q). Economics sees especially strong government use, while psychology, sociology, and political science see relatively strong media use. Arts and humanities (philosophy, art and history, Fig. 1r-t) are relatively under-represented in all three domains.

Specialization in public use further appears at sub-domain levels (Fig. S15). For government, different agencies consume very different scientific research. For example, the U.S. Department of Treasury draws especially on economics and business research, the U.S. Department of Energy draws especially on geology and engineering, and the U.S. Department of Defense draws unusually on history. Different patenting fields further exhibit highly specialized relationships to specific scientific fields. By contrast, in media, while the Washington Post draws unusually heavily on political science research, mainstream media sources in general are more consistent in the fields they report, with especially strong and widespread interest in medicine and psychology.

Overall, these results highlight a large set of specialized relationships between specific domains of public use and specific fields of scientific research. From a public goods perspective, if we think of scientific fields as akin to a series of national parks, we see that each park is embedded in particular communities of public use. Collectively, these parks spread across diverse regions of knowledge and are accessed by diverse segments of the public. A few fields, and especially biology, receive visitors at relatively intense rates from a broad range of public domains – a "Yellowstone Park" of science.

Our second set of results examine whether the public domains tend to consume ideas that scientists themselves consider impactful. Longstanding arguments suggest that the public is not well equipped to evaluate science and may draw on poorly established scientific ideas, which would undermine the public good benefits of science[6-8]. Continuing the national parks metaphor, scientists may be seen as primarily working in a hard-to-reach backcountry, with the typical public visitor having neither the tools to access this terrain nor perhaps an interest in the areas the scientists focus upon. To further examine public use, we therefore consider, at the article level, the alignment between public use and scientific use. Specifically, we calculate the probability of being a hit paper within science, defined as those papers in the top 1% of citations within the same field and year, and examine the relationship to usage in the public domains (SI S4.1, Fig. 2b). We find that papers referenced in public domains have a remarkably high likelihood of being hit papers within science. Papers cited by government documents, news or patents exhibit hit rates of 16.3%, 18.5% and 10.5%, respectively, all large multiples of the baseline rate of 1%. Further, papers referenced in the intersection of different domains tend to be exceptionally impactful in science. For papers referenced in two public domains, approximately half are hit papers. Papers referenced by both government documents and news media have a hit rate of 51.7%. The results are broadly similar if we examine the intersection between government documents and patents (46.0%) or news and patents (52.1%). Lastly, a paper consumed in all three domains is a hit paper in science at a staggering 82.8 times the baseline rate.

The use of high-impact papers is not only common across different public domains, it also appears universal across research areas. Papers covered by public domains tend to be highly cited in all scientific fields, as these papers' hit rates all exceed the baseline by large multiples in each field (Fig. 2c-e). These findings remain similar when varying the threshold for hit papers to top 5% or 10% citations (SI S5.3, Fig. S10,11). We also repeated our analyses for papers produced by U.S.-based researchers, arriving at the same conclusions (Fig. S12). While government, media, and patenting documents may cite science for a variety of reasons and our reference-based measures are proxies for uses of science[26,37,38], we see that the science referenced in public domains is not in conflict with what scientists themselves consider important; rather, impactful papers defined by these communities show substantial overlap. This finding stands in contrast to concerns that the government and media in particular may be poorly positioned to distinguish between high and low impact scientific work[6-8,29,39]. Overall, the public use of science, while marked by substantial specialization in use across research areas, presents a striking universality, where diverse public domains all draw on the highest-impact scientific papers within each field.

We further fine-grain the 19 broad research fields of papers into 294 subfields as indexed by MAG, and calculate the *RCI* score for each subfield in a given public domain. We visualize each field's *RCI* values, locating each field within a common triangle to compare each field's tendency toward usage in specific public domains (Fig. 3a). Fields in social science as well as arts and humanities are mostly used in media and government, whereas fields in science and engineering spread out widely within the triangle, again highlighting the field-level specialization yet collective diversity in the public uses of science.

Together, these results raise a central question: To what degree does the funding input for science relate to the field's public use? The majority of scientific research is supported by public investment, which aims to advance not only science itself but also broader public interest[38]. The NSF, for example, formally introduced broader impacts as a key criterion for evaluating grant proposals in 1997. Here we focus on U.S.-funded projects and use $D_5$ to calculate the average funding per paper in a given subfield as a proxy for public investment costs per unit of output.

We find that the public investment per paper differs dramatically across fields, spanning over five orders of magnitude. Yet comparing average funding per paper with *RCI* in each domain reveals substantial correlations between funding and the use of science across all three public domains, with $R^2 = 0.202$ for government, 0.318 for news, and 0.373 for patents (Fig. 3b-d, SI S4.2, Table S1). To further test if the uncovered correlation is due to the heterogeneity in field size or parent field, we add the number of papers in the subfield as well as parent field fixed effects (for the 19 higher-level fields) into the regression, finding the strong correlation with *RCI* persists ($P < 0.001$ in all three cases). Although a majority of funding come from government agencies, the correlation with funding is the weakest for government documents. Indeed, across the three domains, the representation of subfields in government documents has the lowest predictive power for funding, suggesting that public investments in science better reflect the overall public interest captured by media or patents. We further include funding from non-governmental sources, finding our conclusions remain the same (SI S5.1, Fig. S7, Table S4-5).

Most strikingly, a simple linear regression model combining the three *RCI* values together yields a surprisingly high degree of agreement with funding, with an $R^2$ of 0.672 (Fig. 3e, SI S4.2, Table S2), providing at minimum an 80% increase in predictive power compared with using any of the three public domains alone. These results suggest that each public domain provides independent predictive power for understanding the allocation of public investment in science. The uncovered high predictive power of this analysis is especially striking given many complex factors and processes at work in appropriations, budget setting, and grant review[40-43]. Although each research field differs significantly in its relative role and contribution in science and beyond, the combination of their impacts beyond science powerfully predicts funding, suggesting that, ultimately, what the public uses, what scientists use, and what is funded are remarkably consistent.

Taken together, the analysis probes quantitatively key features of the public good approach to science and establishes systematic facts. Measuring the usage of scientific research outside science itself, we uncover enormous diversity and specialization in how different fields of scientific inquiry are linked to different public domains. Yet, despite these differences, the different public domains (and subdomains) universally draw on highly cited papers within science, indicating that public use is strongly aligned with what scientists themselves consider impactful. And, critically, the public usage of scientific fields across the diverse domains provides simple yet powerful predictors for the level of public investment in each field.

Note that, although the three domains each represent an important dimension of the public space, they do not cover all domains that science may impact. Even within each of the three domains we studied, there may be consumption of science through channels that go beyond our datasets. For example, scientists and their ideas can appear through television, in congressional testimony, and in private sector consulting. Nevertheless, our datasets allow us to quantitatively examine public uses of science across all scientific fields by diverse public domains, revealing individually specialized and collectively diverse uses, universality in impact, and a remarkable alignment between the funding of science and its public use.

As society's support of science depends on a public goods model[21,23], and as legislators have called for more transparency in the usage and value of scientific funding[44], the framework developed in this paper provides an empirical tool, offering quantitative evidence to inform discussions around public good features of science. The allocation of science funding involves chains of decisions by individuals and groups with different perspectives and priorities. These considerations range from legislative committees and the goals of individual political representatives, to funding agency leaders, to within-agency mechanisms that often incorporate insights from scientists, interacting in a complex process that must bridge across distinct communities. As such, one might expect a substantial disconnect between what is eventually funded and forms of public interest – metaphorically, funding of public parks in ways weakly related to public use. Yet, despite the massive diversity in the public uses of science and a complex funding process, there is remarkable alignment in the end result. What the public uses and what scientists themselves use are closely consistent. And the funding of science closely tracks quantifiable public use. These results suggest the connections between the ivory tower and the real world appear more aligned than is commonly imagined.

1 Jewkes, J. *The sources of invention*. (Springer, 1969).

2 Gibbons, M. & Johnston, R. The roles of science in technological innovation. *Research policy* **3**, 220-242 (1974).

3 Landau, R., Rosenberg, N. & National Academy of Engineering. *The positive sum strategy: Harnessing technology for economic growth*. (National Academies Press, 1986).

4 Mansfield, E. Academic research and industrial innovation. *Research policy* **20**, 1-12 (1991).

5 Klevorick, A. K., Levin, R. C., Nelson, R. R. & Winter, S. G. On the sources and significance of interindustry differences in technological opportunities. *Research policy* **24**, 185-205 (1995).

6 Landry, R., Lamari, M. & Amara, N. The extent and determinants of the utilization of university research in government agencies. *Public Administration Review* **63**, 192-205 (2003).

7 Dunn, W. N. The two-communities metaphor and models of knowledge use: An exploratory case survey. *Knowledge* **1**, 515-536 (1980).

8 Snow, C. P. *Science and Government*. (Harvard University Press, 2013).

9 Hatfield, E. Proxmire's golden fleece award. *Relationship Research News (Newsletter of the International Association for Relationship Research)* **4**, 5-9 (2006).

10 Coburn, T. *The National Science Foundation: Under the microscope*. (Senator Tom Coburn, 2011).

11 Ridley, M. *The evolution of everything: How new ideas emerge*. (HarperCollins, 2015).

12 Kealey, T. The case against public science. *Cato Unbound* **5** (2013).

13 Disraeli, B. *Inaugural address delivered to the University of Glasgow Nov. 19, 1873*. (Longmans, Green, and Co., 1873).

14 Merton, R. K. *The sociology of science: Theoretical and empirical investigations*. (University of Chicago press, 1973).

15 Gibbons, M. *The new production of knowledge: The dynamics of science and research in contemporary societies*. (Sage, 1994).

16 Mokyr, J. *The gifts of Athena: Historical origins of the knowledge economy*. (Princeton University Press, 2002).

17 Etzkowitz, H. & Leydesdorff, L. The dynamics of innovation: from National Systems and "Mode 2" to a Triple Helix of university–industry–government relations. *Research policy* **29**, 109-123 (2000).

18 Committee on Prospering in the Global Economy of the 21st Century, National Academy of Sciences & National Academy of Engineering Institute of Medicine. *Rising above the gathering storm: Energizing and employing America for a brighter economic future*. (National Academies Press, 2014).

19 Hjort, J., Moreira, D., Rao, G. & Santini, J. F. How research affects policy: Experimental evidence from 2,150 brazilian municipalities. Report No. 0898-2937, (National Bureau of Economic Research, 2019).

20 Jefferson, T. No patent on ideas: letter to Isaac McPherson. *August* **13**, 1813 (1813).

21 Arrow, K. Economic welfare and the allocation of resources for invention. The rate and direction of inventive activity: economic and social factors. *N. Bureau* (1962).

22 Stiglitz, J. E. Knowledge as a global public good. *Global public goods* **1**, 308-326 (1999).

23 Stephan, P. E. The economics of science. *Journal of Economic literature* **34**, 1199-1235 (1996).

24 Caplan, N. The two-communities theory and knowledge utilization. *American behavioral scientist* **22**, 459-470 (1979).

25 Lynn, L. E. Knowledge and policy: The uncertain connection. (1978).

26 National Research Council. *Using science as evidence in public policy*. (National Academies Press, 2012).

27 Langrish, J., Gibbons, M., Evans, W. G. & Jevons, F. R. *Wealth from knowledge: Studies of innovation in industry*. (Springer, 1972).

28 Yin, Y., Gao, J., Jones, B. F. & Wang, D. Coevolution of policy and science during the pandemic. *Science* **371**, 128-130 (2021).

29 Thelwall, M., Haustein, S., Larivière, V. & Sugimoto, C. R. Do altmetrics work? Twitter and ten other social web services. *PloS one* **8** (2013).

30 Costas, R., Zahedi, Z. & Wouters, P. Do "altmetrics" correlate with citations? Extensive comparison of altmetric indicators with citations from a multidisciplinary perspective. *Journal of the Association for Information Science and Technology* **66**, 2003-2019 (2015).

31 Ahmadpoor, M. & Jones, B. F. The dual frontier: Patented inventions and prior scientific advance. *Science* **357**, 583-587 (2017).

32 Fleming, L., Greene, H., Li, G., Marx, M. & Yao, D. Government-funded research increasingly fuels innovation. *Science* **364**, 1139-1141 (2019).

33 Marx, M. & Fuegi, A. Reliance on science: Worldwide front-page patent citations to scientific articles. *Strategic Management Journal* (2020).

34 Wang, K., Shen, Z., Huang, C., Wu, C.-H., Dong, Y. & Kanakia, A. Microsoft Academic Graph: When experts are not enough. *Quantitative Science Studies* **1**, 396-413 (2020).

35 Kosack, S., Coscia, M., Smith, E., Albrecht, K., Barabási, A.-L. & Hausmann, R. Functional structures of US state governments. *Proceedings of the National Academy of Sciences* **115**, 11748-11753 (2018).

36 Herzog, C., Hook, D. & Konkiel, S. Dimensions: Bringing down barriers between scientometricians and data. *Quantitative Science Studies* **1**, 387-395 (2020).

37 Weiss, C. H. The many meanings of research utilization. *Public Administration Review* **39**, 426-431 (1979).

38 Bornmann, L. What is societal impact of research and how can it be assessed? A literature survey. *Journal of the American Society for Information Science and Technology* **64**, 217-233 (2013).

39 Selvaraj, S., Borkar, D. S. & Prasad, V. Media coverage of medical journals: do the best articles make the news? *PLoS One* **9** (2014).

40 Boudreau, K. J., Guinan, E. C., Lakhani, K. R. & Riedl, C. Looking across and looking beyond the knowledge frontier: Intellectual distance, novelty, and resource allocation in science. *Management Science* **62**, 2765-2783 (2016).

41 Bromham, L., Dinnage, R. & Hua, X. Interdisciplinary research has consistently lower funding success. *Nature* **534**, 684 (2016).

42 Ginther, D. K., Schaffer, W. T., Schnell, J., Masimore, B., Liu, F., Haak, L. L. & Kington, R. Race, ethnicity, and NIH research awards. *Science* **333**, 1015-1019 (2011).

43 Ma, A., Mondragón, R. J. & Latora, V. Anatomy of funded research in science. *Proceedings of the National Academy of Sciences* **112**, 14760-14765 (2015).

44 Hearing: The science of science and innovation policy. *Committee on Science and Technology* (2010).

**Acknowledgements** We thank Iris Shen, Darrin Eide, and all members of Microsoft Academic group for their invaluable help. This work uses data sourced from Altmetric.com and Dimensions.ai through researcher access plans and is supported by the Air Force Office of Scientific Research under award number FA9550-17-1-0089 and FA9550-19-1-0354, National Science Foundation grant SBE 1829344, the Alfred P. Sloan Foundation G-2019-12485.

**Author contributions** D.W., B.F.J. and K.W. conceived the project and designed the experiments; Y.Y. and Y.D. collected data; Y.Y. performed empirical analyses with help from D.W.,B.F.J., Y.D. and K.W.; all authors discussed and interpreted results; Y.Y., B.F.J and D.W. wrote the manuscript, all authors edited the manuscript.

**Competing interests** Y.D. and K.W. are employees of Microsoft Corporation. Y.Y., B.F.J. and D.W. declare no competing interests.

**Data availability** Deidentified data necessary to reproduce all plots and statistical analyses will be made freely available. MAG and paper-patent linkage data are publicly available. Those who are interested in raw data of Altmetric and Dimensions should contact Digital Science directly.

**Code availability** Code will be made freely available.

**Figure captions**

**Fig. 1. Diversity in public use.** Different scientific fields experience distinct and typically specialized public uses. (**a-t**) The usage metric *RCI* for the three public domains, presented for each field (**b-t**). The dashed triangles represent a null model where each paper has the same chance to be used (**a**).

**Fig. 2. Public use and scientific use.** The public tends to consume exceptionally high impact science from all fields and in all three public domains, indicating alignment between public use and scientific use. (**a**) Usage by domain for papers published from 2005 to 2014. The area of each subset is proportional to the square root of the paper count in the corresponding public domain. (**b**) Hit rates for papers cited in one, two, or three public domains. Hit papers are defined as those receiving citation counts, within science, in the top 1% within the field and year. (**c-e**) Hit rates for each of the 19 fields consumed by government documents (**c**), news media (**d**) and patents (**e**). In all fields, and in all three domains, the consumed papers have hit rates within science many times larger than the baseline rate of 1% (dashed line).

**Fig. 3. Public use and public funding.** Amidst enormous diversity in public use across fields and domains, scientific funding for a given field is closely aligned with the totality of its public use. (**a**) Ternary plot of *RCI* for 294 level-1 fields together, with the location of each field indicating its relative usage among the public domains. Circles are colored coded according to its parent field in Fig. 1, and circle sizes reflect overall usage. (**b-d**) Average funding per paper across fields is positively correlated with a field's *RCI* index in government (**b**), news (**c**) and patenting (**d**). The relationship remains significant when combined with control variables ($P <$ 0.001 in all three cases after controlling for the number of papers and parent field fixed effects). (**e**) Collectively, public uses beyond science strongly predict field level funding per paper.

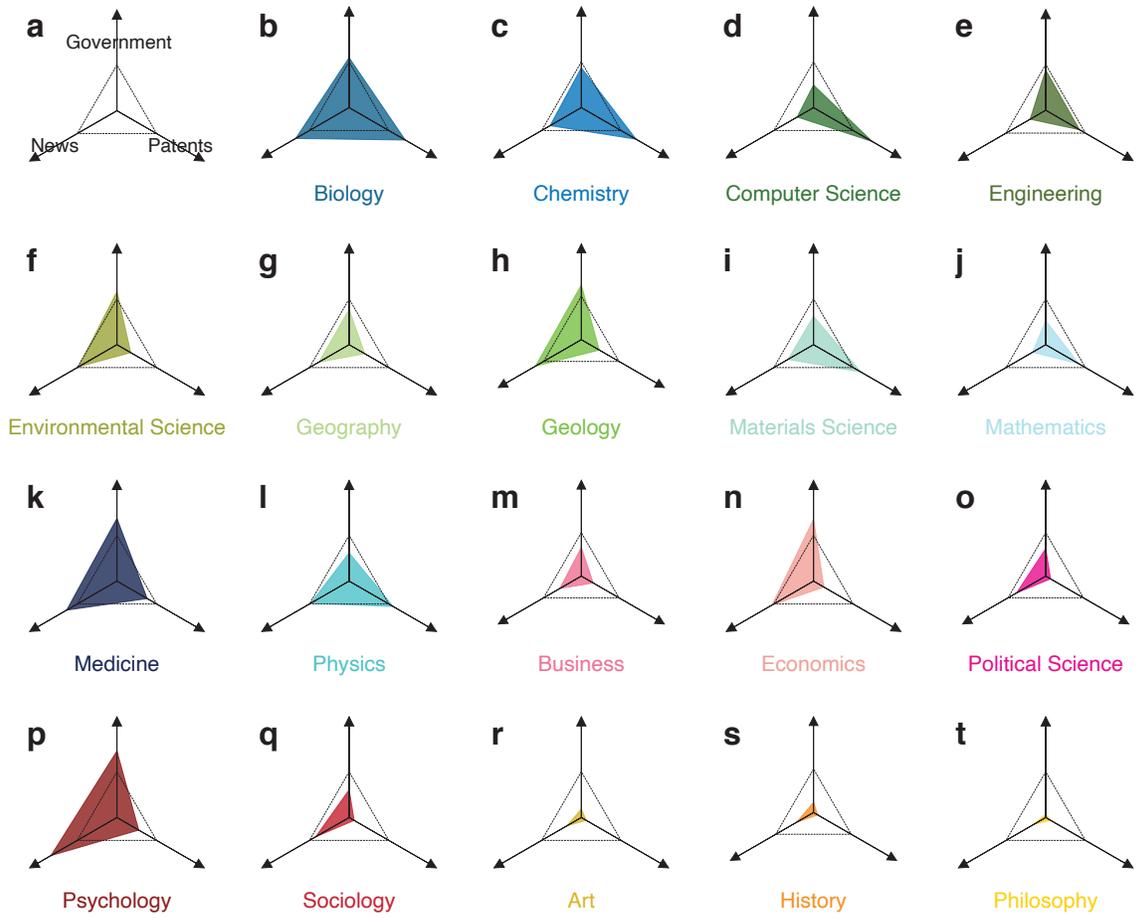

**Fig. 1. Diversity in public use.**

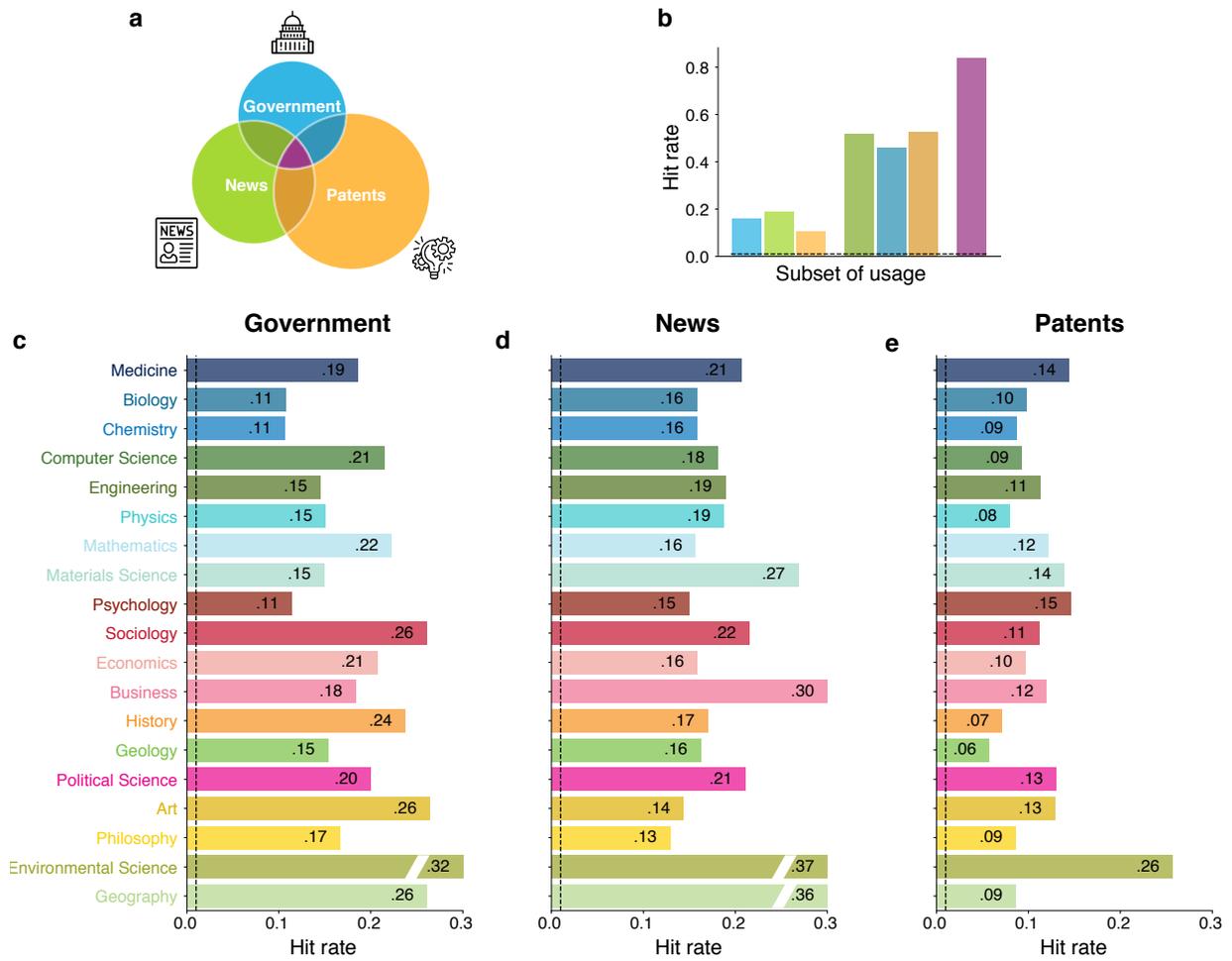

**Fig. 2. Public use and scientific use.**

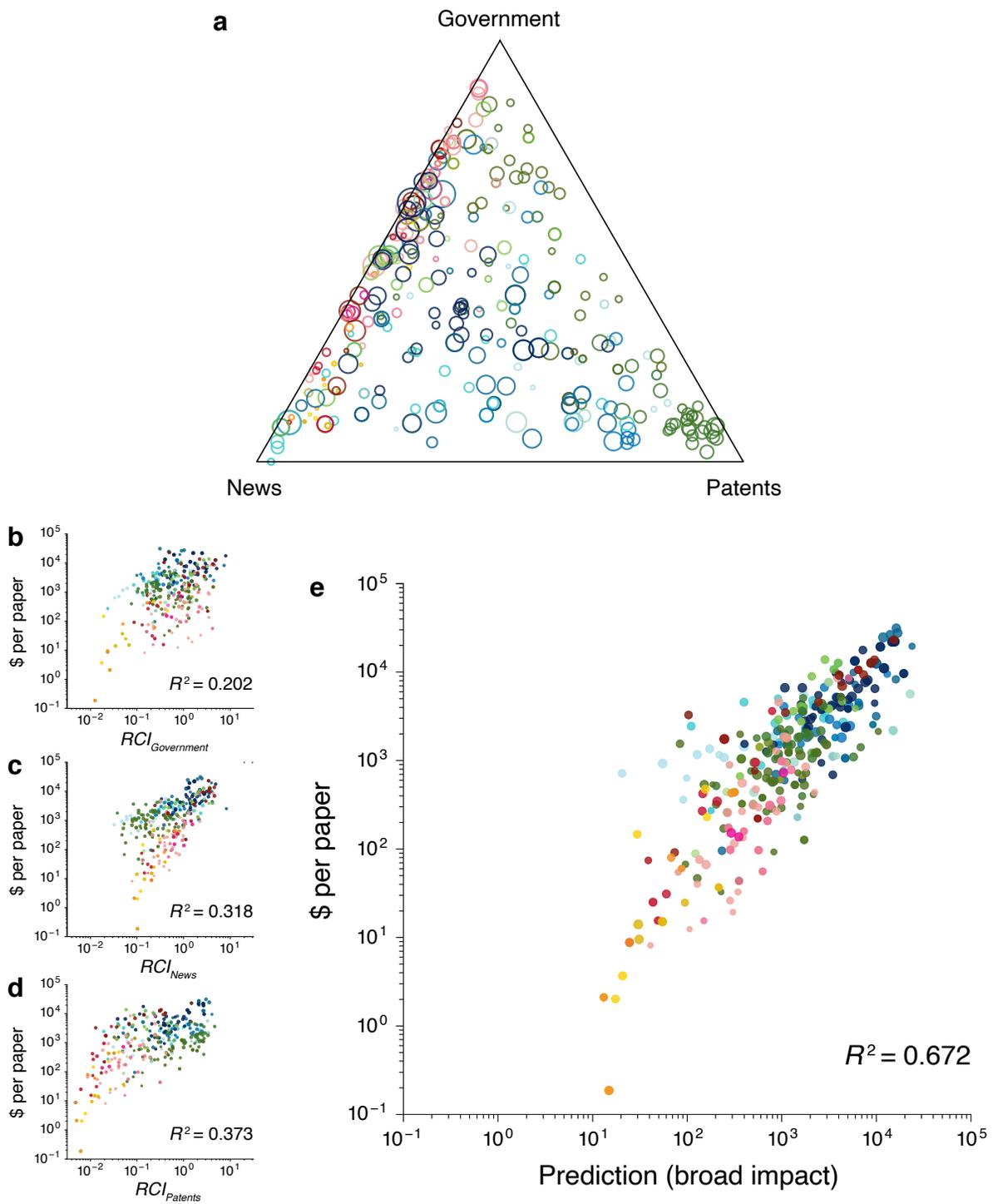

**Fig. 3. Public use and public funding.**

# Supplementary Information for
# Science as a Public Good: Public Use and Funding of Science

Yian Yin[1,2,3], Yuxiao Dong[4], Kuansan Wang[4], Dashun Wang[1,2,3,5]*, Benjamin F. Jones[1,2,5,6]*

*Correspondence to dashun.wang@northwestern.edu, bjones@kellogg.northwestern.edu

## Table of Contents

# S1 Data description

**S1.1 Microsoft Academic Graph ($D_1$)**

The publication and citation data are primarily obtained from Microsoft Academic Graph (MAG, accessed Oct 2018)[34,45]. MAG is arguably the largest open-source citation database to date and contains records of 209M documents. We inter-linked different data tables to obtain the author, affiliation, year, venue and topic information for each paper. MAG includes a variety of document types. To focus on scientific articles, we consider publications under the categories of journal papers, conferences papers, books and book chapters, and papers with DOI information. In other words, two kinds of publications are excluded in our analysis: patents and papers with neither category nor DOI information. We further focus on papers published in a 10-year period from 2005 to 2014, leading to a subset of 36M papers in total.

MAG uses a non-mutually exclusive hierarchy for research topic (field of study) mined by semantic analysis tools. To explore major research fields, the analysis considers the level-0 (19 fields) and level-1 (294 fields) categorizations (Fig. S1a). Figure S1b and S1c show the number of level-0 and level-1 fields that each paper is connected to, indicating both quantities are narrowly distributed.

**S1.2 US Government Documents ($D_2$)**

To quantify references to scientific articles in the government domain, one needs to construct a large-scale dataset of government documents that can be linked to the scientific papers. The task has been difficult in part because government documents are spread across many sources. Despite recent efforts like govinfo[46] to digitalize and standardize government publication information, most existing sources have relatively low coverage, especially for the executive branch. Furthermore, although a significant fraction of such documents may cite scientific literature, such citations do not follow a common structure. To tackle both challenges, here we develop a novel pipeline to construct the dataset.

Our data collection starts with a list of URLs under the .gov domain, which is the domain name for government agencies and contains the vast majority of U.S. government entities. Given the huge number of such pages, here we use a PageRank-like assessment system provided by the Microsoft Bing search engine, which assigns each URL with a tiered index of importance. In this study, we focus on Tier 0 (the most important) pages to construct the sample, which contains approximately 6M URLs within the .gov domain. We downloaded these pages using an automatic crawler and focused on all PDF files in this set (~28% among the corpus). We also notice that a small proportion of these documents are themselves research papers, as their urls can be linked to MAG papers though MAG paper url table, and exclude these documents from our analysis.

To extract the references cited in these files we apply Science-Parse[47], an open-source tool for reference string extraction developed by the Allen Institute for Artificial Intelligence. Science-Parse is a state-of-art framework that scans PDF files and returns a list of all reference-like strings. We then match this list to the MAG. Since the PDF reference extraction may contain minor errors, exact title matching of paper items may not be the optimal approach. Specifically, we indexed the full MAG to compile a search engine-like system using title, journal, author and publication year information. By leveraging the Okapi BM25 measure[48], one of the core algorithms used by modern full-text search engines, we query each string to obtain a list of the top 2 candidate paper items, each accompanied by a score representing the degree of agreement. To find the score threshold for determining if a string is successfully matched, here we use scores of the $2^{nd}$ matched paper as a null model for score distributions (Fig. S2a). Indeed, assuming the score is a reasonable quantification, one would expect the difference of $1^{st}$ and $2^{nd}$ matched paper to be significant if and only if the $1^{st}$ ranked paper is a true match of the string. To this end, for each string we first calculate the score distribution of all $2^{nd}$ matched scores of the similar query word length as a baseline. The string is considered to be matched to the $1^{st}$ ranked paper when the score is significantly higher than a right tail cutoff of the baseline distribution ($P = 0.05$). We further test this algorithm by comparing its predictions with manual validations on 100 randomly selected papers through two evaluations: (1) For a binary classification based on whether a reference string can be matched to a MAG paper, we calculate the F1 score; and (2) conditional on being classified as positive in (1), we measure the accuracy of the matched MAG document ID. We find high consistency between the results returned by the algorithm and our manual validations (Fig. S2b).

### S1.3 Altmetric dataset ($D_3$)

To study references to scientific publications in the news media, we use a dataset offered by Altmetric[29,30,49]. This dataset records approximately 19.4M papers with at least one news media or social media mention. We then merge paper information with MAG. A vast majority (16.1M) of publications in the Altmetric database have unique digital object identifiers (DOI), allowing us to connect this with DOI information in MAG. We find that 13.7M (84%) of the DOIs can be matched to records in MAG.

### S1.4 USPTO patent database ($D_4$)

To study references to scientific publications in patents, we build on prior work and use a comprehensive list of mapping from United States Patent and Trademark Office (USPTO) patents to MAG papers, which includes approximately 15M citation pairs between patents and papers[50]. To classify patents into technology classes, we use the Cooperative Patent Classification (CPC) system, drawn from PatentsView, a data platform based on USPTO bulk data[51]. Combining the two files provides technology class information for 97.5% of patents that reference scientific articles. The small share of missing technology class cases corresponds to patents granted in late 2018, which have not been updated in our data.

### S1.5 Dimensions scientific funding data ($D_5$)

To understand how research funding from various sources is allocated into different scientific fields, we leverage research funding data from Dimensions[36,52], a novel dataset that includes approximately 5M research projects supported by over 400 funding agencies worldwide. To be consistent with our publication analysis, we focus on projects funded during the same ten-year period (2005-2014). One challenge in our estimation here is that some projects are supported both within and outside the ten-year period, for example, in year 2014, 2015 and 2016, while only the total funding amount is available. Here we estimate funding amount in the ten-period by multiplying the total amount with the fraction of time within the ten-year period, which equals to 1/3 in this example. We further focus on projects that have funding amount information and are funded by US agencies.

A unique opportunity provided by Dimensions is a linkage table between supporting grants and resulting publications, which allows us to categorize the field of each grant according to its resulting publications. More than 90% of the publications have DOI information which can be further matched to items in our paper database. Here we use this table and focus on projects that can be linked to at least one MAG paper (which allows us to estimate the research field for over 74% of U.S. research funding in this period). We then create a list of level-1 fields by combining the fields of resulting publications supported by each grant, and evenly split funding amount of this project to each field in this list. For example, if a grant of $15,000 supports three publications, two in quantum physics and one in mathematical physics respectively, we assign $10,000 to quantum physics and $5,000 to mathematical physics. Together we link 292,875 funded projects with at least one publication.

**S1.6 Data limitations**

Our data are not without limitations. First, our datasets only represent a subset of all possible government documents and media news in the world, and there could be heterogeneity within documents published by different agencies or news covered by different media. Second, the linkage strategy between science and public uses and funding is based on automatic algorithms and may contain some errors. While our validations in S2 and robustness checks in S5 have not uncovered any potential biases, readers should keep in mind of the existence of these factors. Nevertheless, despite these potential limitations, it is important to note that these data sources are among the largest in their respective domains, and approaches for data linkages are also among the most advanced of their kind, hence representing the state-of-art empirical basis to understand the interaction between science and public domains.

# S2 Independent data validation

**S2.1 Overton policy documents ($D_6$)**

Policy documents extracted from the Bing search engine and their associated references provides a novel data that is only possible with recent advances in information retrieval and machine learning. Given the novelty of such applications (reference parsing in policy documents), we lack systematic baseline methods for comparison. Here we leverage another novel dataset, Overton, which has just become available during the writing of this manuscript and provides an independent validation case. Overton is among the world's largest searchable index of policy documents, including about 3M policy-related documents from thousands of sources in different types (government agencies, think tanks and intergovernmental organizations). Overton also extracts scientific references in documents and maps them into DOIs. Here we retrieve all policy documents published by U.S. governmental agencies indexed by Overton, looking at all scientific references (published in the same ten-year period) they have ever cited, and use the DOI to connect these papers into MAG. For comparison we calculate two fundamental measurements that are key to our main conclusions: the relative usage ($RCI$) and average hit rate[28]. In Fig. S3a, we compare the relative usage $RCI$ for level-1 fields, finding high consistency between the two sources (Spearman correlation $r_s = 0.94$, $P < 0.001$). We further calculate paper hit rates used in policy for level-0 fields across Overton and our dataset (Fig. S3b), again finding high correlations (Spearman correlation $r_s = 0.60$, $P = 0.006$). Together, these results suggest that both dimensions studied in this paper seem reliable across different datasets. Given that these two sources are built and developed independently, the remarkable consistency offers strong support for the robustness of our findings.

We further use Overton data to repeat our results in Fig. 3 (Fig. S6). Although the correlation between log of $RCI_{Government}$ and log of average funding decreases ($R^2$ from 0.202 to 0.097, Fig.S6A), the relationship remains significant ($P < 0.001$ after controlling for number of papers and level-0 parent field fixed effect, Table S3).

## S2.2 RePORTER funding dataset ($D_7$)

The Dimensions data ($D_5$) is a state-of-art database linking funding and associated publications, using information from funding agencies as well as text mining from the acknowledgement section of publications. As an alternative, we further leverage funding-paper linkage information from two major funding sources in the U.S. – the National Science Foundation (NSF) and National Institutes of Health (NIH). These two agencies are the largest federal funders for scientific research and together account for more than half of overall federal research funding[53]. Information on all projects funded by NIH over the last several decades and resulting publications are available through NIH RePORT (Research Portfolio Online Reporting Tools), an open data source developed since 2008[54]. Bulk data for NSF grants in a similar format are available as part of Federal RePORTER, a federal effort to "create a repository of data and tools that will be useful to assess the impact of federal R&D investments"[55].

To test our data coverage, we calculated the number of publications supported by each grant active in the ten-year period (2005-2014) from both RePORTER and Dimensions data. We find in both NSF and NIH, the number of publications supported by each grant is highly correlated, showing a high degree of consistency between the two data sources (Fig. S4bd). Further, we find that Dimensions reports more resulting publications on average (Fig. S4ac). The superiority of Dimensions may be explained by multiple reasons, including an incomplete coverage of data in early years from Federal RePORTER and the fact that more complete paper-grant linkages can be found through publication acknowledgements. Regardless, these results suggest Dimensions is the preferred source for linking papers and grants for this study.

## S3 Related works

### S3.1 Policy uses of science

There has been a long-lasting interest in understanding uses of science in policy and decision-making over many decades[8,26], offering a rich set of conceptual models capturing policy uses of science, ranging from two communities theory[7,24] that suggest little common language and interest between scientists and policymakers, to supply-side and demand-side models that highlight knowledge production on the science side[56] or problem solving on the policy side[37] as the primary driver of research utilization in policy, to interaction models that depicts an iterative process where science and policy co-evolve[26,57,58]. Some works along this line has also examined different types of policy uses, arguing that policy-science citations may occur for various reasons[59,60], including (i) instrumental uses (knowledge directly applied to solve problems); (ii) conceptual uses (research that influences or informs the way policymakers think); (iii) tactical uses (citing research to support or challenge an idea) among others, suggesting value in understanding the semantics of the policy-science citations. Yet at the same time, it has also been recognized that distinguishing these uses at scale remains a challenging task[26]. Here we develop a scaled approach based on citation links and leave analysis of how specifically the science is being used to further investigation.

**S3.2 Altmetrics literature**

Altmetrics studies alternative or complementary indicators related to scientific publications[29,30,61]. This field leverages databases such as Altmetric and PlumX and has grown rapidly in the last decade, deepening our empirical understanding[61] of how science is covered across different online platforms, including social media platforms, mainstream media news, policy documents, Wikipedia and other sources. For example, several prior studies in this literature have examined the overall coverage and reliability of such datasets[62-64]. Based on these datasets, researchers have also examined how scientific papers from different fields receive attention beyond citations[65], as well as the relationship between citation metrics and altmetrics indexes[29,30]. For a more detailed overview of the Altmetrics literature, readers may refer to two recent reviews[61,66].

**S3.3 Scientific non-patent references**

Scientific non-patent references have also been studied in the recent innovation literature, partly due to the increasing availability of approaches and data sources for large-scale matching between patent references and scientific publications. Recent systematic linkage efforts have connected USPTO patents to Web of Science[67] and Microsoft Academic Graph[33]. Existing literature has examined meanings of such citations, suggesting these linkages as a useful signal for association between science and technology[68,69]. Furthermore, scientific non-patent references have been leveraged to construct higher-order indirect links between patenting and science[31,32].

# S4 Methodology

**S4.1 Citation percentiles and hit papers**

Despite the widespread use of citations as a proxy for scientific impact, direct comparison of citation counts received by papers across time and field can be problematic without normalization[70]. We therefore calculate citation percentiles for papers within the same publication year and field. Here following prior studies[31,71,72], we define 'hit papers' (also known as 'home runs') as papers ranking in the top 1% of citations received. As we tune the threshold from 1% to 5% or 10%, we find the result that high-impact papers are disproportionately used by public domains remains robust (See SM S5.3, Fig. S10,11).

**S4.2 Regression models**

To understand the association between public use and funding for different scientific fields, we use linear regression models (ordinary least squares). We first note that all three *RCI* measures are highly skewed (Fig. S5a-c), prompting us we take the natural logarithm (Fig. S5d-f), ln *RCI*, in our linear regressions. The same transformation is taken on the average funding per paper.

The variables are defined as follows:
*Dependent variable*: The dependent variable is $\ln Y_i$, defined as the natural logarithm of average funding per paper.
*Predictors of interest*: We examine in regression the extent to which different impact measures can predict funding, including $\ln RCI_j$ for the three public domains, as well as $\ln c$, the natural logarithm of mean citations received for papers in that field. To include all data points in the regression, for the rare cases when an impact measure is 0, we add 1 to avoid 0-s in the logarithm. We further include the natural logarithm of the number of papers published in the ten-year period, $\ln p$, as another control variable.
*Fixed effects*: To control for the possible fact that fields under different broad categories may have specific way of funding and public use and hence are not directly comparable, we introduce $F_f$, fixed effect terms for each level-0 field. Specifically, $F_{fi} = 1$ if the level-1 field *i* is a child field of the level-0 field *f* according to MAG's classification structure. Note that some level-1 fields belong to two level-0 fields simultaneously (e.g. mathematical physics is the child field of both mathematics and physics).

We start with bivariate regressions examining the relationship between each *RCI* (i.e. for government, media, or patenting) and average funding (Table S1, Models 1-3). That is,
$$\ln Y_i = \beta_j \ln RCI_{ji} + \varepsilon_i$$

where $i$ indexes a specific level-1 field, $j$ indexes one of the three public domains, and $Y_i$ is average funding per paper in the field.

In multivariate regressions, we further include control variables to test if the uncovered correlation is simply due to the heterogeneity in field size or parent field (Table S2, Models 4-6)

$$\ln Y_i = \beta_j \ln RCI_{ji} + \beta_p \ln p_i + \sum_f \beta_f F_{fi} + \varepsilon_i$$

Together these results indicate that funding per paper across fields is strongly positively correlated with relative usage in any of the three domains when analyzed separately. This finding prompts us to investigate the joint predictive power of the three *RCI*s (Table S2, Model 7):

$$\ln Y_i = \sum_j \beta_j \ln RCI_{ji} + \varepsilon_i$$

which shows that each measure contributes independently and substantially to explaining the variation in funding.

Finally, we add further control variables into Models 7 (Table S2, Model 8):

$$\ln Y_i = \sum_j \beta_j \ln RCI_{ji} + \beta_p \ln p_i + \sum_f \beta_f F_{fi} + \varepsilon_i$$

## S5 Further robustness checks

### S5.1 Including non-governmental funding

In our main regression analysis, we mainly focused on US governmental funding agencies (defined as funding agencies under .gov or .mil domains.) Here we run further robustness checks by considering other US funding agencies. Indeed, we find funding information recorded in Dimensions is primarily dominated by governmental funding agencies, and our results remain robust when considering all US funding.

The relationship documented in Fig. 3b-d remains significant (Fig. S7a-c) and significance remains in all three cases after controlling for number of papers and level-0 parent field fixed effect in ($P < 0.001$, Table S4). We also find a similar level of predictive power, where the three public use variables have $R^2 = 0.672$ (Fig. S7d, Table S5).

**S5.2 Changing the criterion of academic publications**

As mentioned in S1.1, papers with neither document categorization nor DOI information are not considered, since this is a noisier population that may not represent standard research publications. Here we repeat our analysis by including these samples and calculate RCI for each level-0 field, finding our results are broadly consistent with Fig. 1 (Fig. S8). We also try another variant by restricting the samples used in main text to those only published in English, again finding our results do not change (Fig. S9).

**S5.3 Robustness of paper hit rate**

We check our definition of hit papers by changing the threshold to top 5% (Fig. S10) or 10% (Fig. S11) highly cited papers. Both variants show that papers used across public domains and scientific fields are universally impactful within science.

We further repeated our analyses for papers produced by U.S.-based researchers. More specifically, we identify U.S. institutions in MAG based on their GRID (Global Research Identifier Database) id and limit our analysis to papers produced by scholars with these institutional affiliations, again finding universal high impact of papers (Fig. S12).

45 Wang, K., Shen, Z., Huang, C.-Y., Wu, C.-H., Eide, D., Dong, Y., Qian, J., Kanakia, A., Chen, A. & Rogahn, R. A Review of Microsoft Academic Services for Science of Science Studies. *Frontiers in Big Data* **2**, 45 (2019).

46 U.S. Government Publishing Office. *Govinfo*, <https://www.govinfo.gov> (2019).

47 AI2. *Science Parse*, <https://github.com/allenai/science-parse> (2019).

48 Robertson, S. & Zaragoza, H. *The probabilistic relevance framework: BM25 and beyond*. (Now Publishers Inc, 2009).

49 Adie, E. & Roe, W. Altmetric: Enriching scholarly content with article-level discussion and metrics. *Learned Publishing* **26**, 11-17 (2013).

50 Marx, M. & Fuegi, A. Reliance on Science: Worldwide Front-Page Patent Citations to Scientific Articles. *Boston University Questrom School of Business Research Paper* (2019).

51 *PatentsView*, <http://www.patentsview.org> (2019).

52 Hook, D. W., Porter, S. J. & Herzog, C. Dimensions: building context for search and evaluation. *Frontiers in Research Metrics and Analytics* **3**, 23 (2018).

53 American Association for the Advancement of Science. *AAAS Federal R&D Budget Dashboard*, <https://www.aaas.org/programs/r-d-budget-and-policy/federal-rd-budget-dashboard> (2020).

54 *NIH Research Portfolio Online Reporting Tools (RePORT)*, <https://report.nih.gov> (2019).

55 *Federal RePORTER*, <https://federalreporter.nih.gov> (2020).

56 Havelock, R. G. *Planning for innovation through dissemination and utilization of knowledge*. (Center for Research on Utilization of Scientific Knowledge, Institute for …, 1979).

57 Landry, R., Amara, N. & Lamari, M. Climbing the ladder of research utilization: Evidence from social science research. *Science Communication* **22**, 396-422 (2001).

58 Jasanoff, S. *States of knowledge: the co-production of science and the social order*. (Routledge, 2004).

59 Jasanoff, S. S. Contested boundaries in policy-relevant science. *Social Studies of Science* **17**, 195-230 (1987).

60 Hilgartner, S. The dominant view of popularization: Conceptual problems, political uses. *Social Studies of Science* **20**, 519-539 (1990).

61 Tahamtan, I. & Bornmann, L. Altmetrics and societal impact measurements: Match or mismatch? A literature review. *El profesional de la información (EPI)* **29**, e290102 (2020).

62 Ortega, J. L. Blogs and news sources coverage in altmetrics data providers: a comparative analysis by country, language, and subject. *Scientometrics* **122**, 555-572 (2020).

63 Haustein, S. Grand challenges in altmetrics: heterogeneity, data quality and dependencies. *Scientometrics* **108**, 413-423 (2016).

64 Bornmann, L. Do altmetrics point to the broader impact of research? An overview of benefits and disadvantages of altmetrics. *Journal of informetrics* **8**, 895-903 (2014).

65 Bornmann, L. Validity of altmetrics data for measuring societal impact: A study using data from Altmetric and F1000Prime. *Journal of informetrics* **8**, 935-950 (2014).

66 Sugimoto, C. R., Work, S., Larivière, V. & Haustein, S. Scholarly use of social media and altmetrics: A review of the literature. *Journal of the Association for Information Science and Technology* **68**, 2037-2062 (2017).

67 Gaetani, R. & Bergolis, M. L. The economic effects of scientific shocks. *Unpublished Manuscript* (2015).

68 Hicks, D., Breitzman, T., Olivastro, D. & Hamilton, K. The changing composition of innovative activity in the US—a portrait based on patent analysis. *Research policy* **30**, 681-703 (2001).

69 Meyer, M. Does science push technology? Patents citing scientific literature. *Research policy* **29**, 409-434 (2000).

70 Radicchi, F., Fortunato, S. & Castellano, C. Universality of citation distributions: Toward an objective measure of scientific impact. *Proceedings of the National Academy of Sciences* **105**, 17268-17272 (2008).

71 Wang, Y., Jones, B. F. & Wang, D. Early-career setback and future career impact. *Nature communications* **10**, 1-10 (2019).

72 Wuchty, S., Jones, B. F. & Uzzi, B. The increasing dominance of teams in production of knowledge. *Science* **316**, 1036-1039 (2007).

| Model | (1) | (2) | (3) |
|---|---|---|---|
| VARIABLES | | | |
| Policy (RCI) | 0.703*** | | |
| | (0.082) | | |
| News (RCI) | | 0.935*** | |
| | | (0.080) | |
| Patent (RCI) | | | 0.678*** |
| | | | (0.051) |
| Observations | 294 | 294 | 294 |
| R2 | 0.202 | 0.318 | 0.373 |
| F | 73.84 | 136.1 | 173.3 |
| rmse | 1.824 | 1.555 | 1.491 |

Standard errors in parentheses
*** p<0.01, ** p<0.05, * p<0.1

**Table S1: Regression results for Models 1-3 (S4.2).**

| Model | (4) | (5) | (6) | (7) | (8) |
|---|---|---|---|---|---|
| VARIABLES | | | | | |
| Policy (RCI) | 0.538*** | | | 0.175*** | 0.197*** |
| | (0.072) | | | (0.063) | (0.063) |
| News (RCI) | | 0.839*** | | 0.794*** | 0.701*** |
| | | (0.073) | | (0.067) | (0.071) |
| Patent (RCI) | | | 0.490*** | 0.640*** | 0.437*** |
| | | | (0.062) | (0.038) | (0.049) |
| # Paper (p) | 0.235*** | 0.298*** | 0.180*** | | 0.205*** |
| | (0.052) | (0.047) | (0.052) | | (0.041) |
| Level-0 fixed effect | Y | Y | Y | | Y |
| Observations | 294 | 294 | 294 | 294 | 294 |
| R2 | 0.707 | 0.763 | 0.713 | 0.672 | 0.825 |
| F | 31.30 | 41.67 | 32.17 | 198.3 | 55.33 |
| rmse | 1.056 | 0.950 | 1.046 | 1.082 | 0.819 |

Standard errors in parentheses
*** p<0.01, ** p<0.05, * p<0.1

**Table S2: Regression results for Models 4-8 (S4.2).**

| Model | (1) | (4) | (7) |
|---|---|---|---|
| VARIABLES | | | |
| Policy (RCI) | 0.414*** | 0.358*** | 0.090 |
| | (0.074) | (0.065) | (0.056) |
| News (RCI) | | | 0.828*** |
| | | | (0.070) |
| Patent (RCI) | | | 0.661*** |
| | | | (0.038) |
| # Paper (p) | | 0.257*** | |
| | | (0.054) | |
| Level-0 fixed effect | | Y | |
| | | | |
| Observations | 294 | 294 | 294 |
| R2 | 0.097 | 0.682 | 0.667 |
| F | 31.47 | 27.77 | 193.3 |
| rmse | 1.789 | 1.100 | 1.091 |

Standard errors in parentheses
*** $p<0.01$, ** $p<0.05$, * $p<0.1$

**Table S3: Regression results using Overton data (S2.1).**

| Model | (1) | (2) | (3) |
|---|---|---|---|
| VARIABLES | | | |
| Policy (RCI) | 0.694*** | | |
| | (0.082) | | |
| News (RCI) | | 0.933*** | |
| | | (0.080) | |
| Patent (RCI) | | | 0.675*** |
| | | | (0.051) |
| Observations | 294 | 294 | 294 |
| R2 | 0.198 | 0.319 | 0.373 |
| F | 72.28 | 136.8 | 173.3 |
| rmse | 1.679 | 1.548 | 1.486 |

Standard errors in parentheses
*** p<0.01, ** p<0.05, * p<0.1

**Table S4: Regression results for Models 1-3 by considering all US funding (S5.1).**

| Model | (4) | (5) | (6) | (7) | (8) |
|---|---|---|---|---|---|
| VARIABLES | | | | | |
| Policy (RCI) | 0.535*** | | | 0.165*** | 0.192*** |
| | (0.072) | | | (0.063) | (0.063) |
| News (RCI) | | 0.840*** | | 0.798*** | 0.705*** |
| | | (0.073) | | (0.066) | (0.071) |
| Patent (RCI) | | | 0.493*** | 0.637*** | 0.439*** |
| | | | (0.062) | (0.038) | (0.049) |
| # Paper (p) | 0.233*** | 0.297*** | 0.178*** | | 0.204*** |
| | (0.052) | (0.047) | (0.052) | | (0.041) |
| Level-0 fixed effect | Y | Y | Y | | Y |
| Observations | 294 | 294 | 294 | 294 | 294 |
| R2 | 0.703 | 0.760 | 0.710 | 0.672 | 0.823 |
| F | 30.67 | 41.07 | 31.74 | 197.8 | 54.65 |
| rmse | 1.059 | 0.951 | 1.046 | 1.078 | 0.820 |

Standard errors in parentheses
*** p<0.01, ** p<0.05, * p<0.1

**Table S5: Regression results for Models 4-8 by considering all US funding (S5.1).**

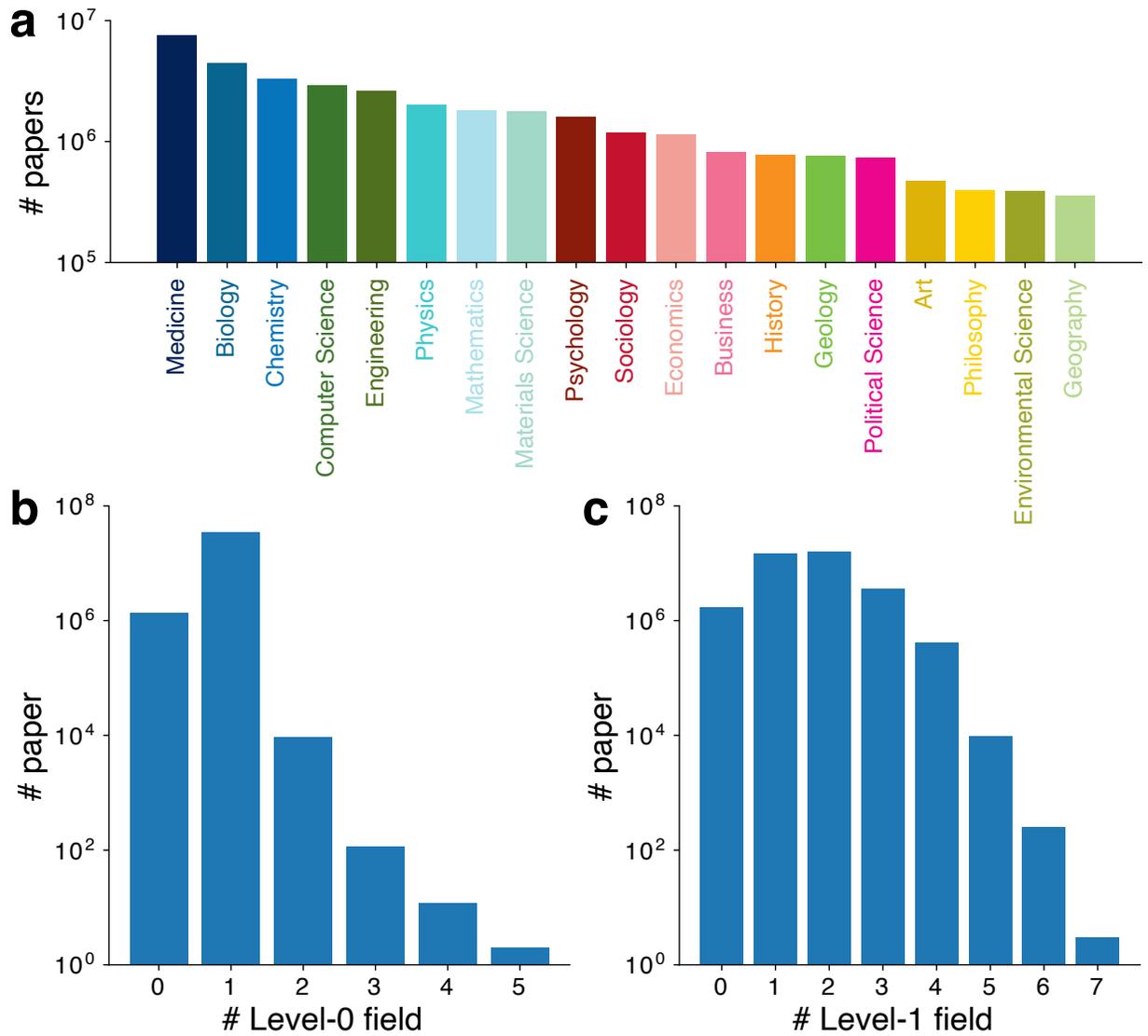

**Figure S1. An overview of MAG field classification.** (**a**) Number of papers belonging to each of the 19 level-0 fields. (**b-c**) Distribution of field numbers that connect to a paper at level 0 (**b**) and 1 (**c**).

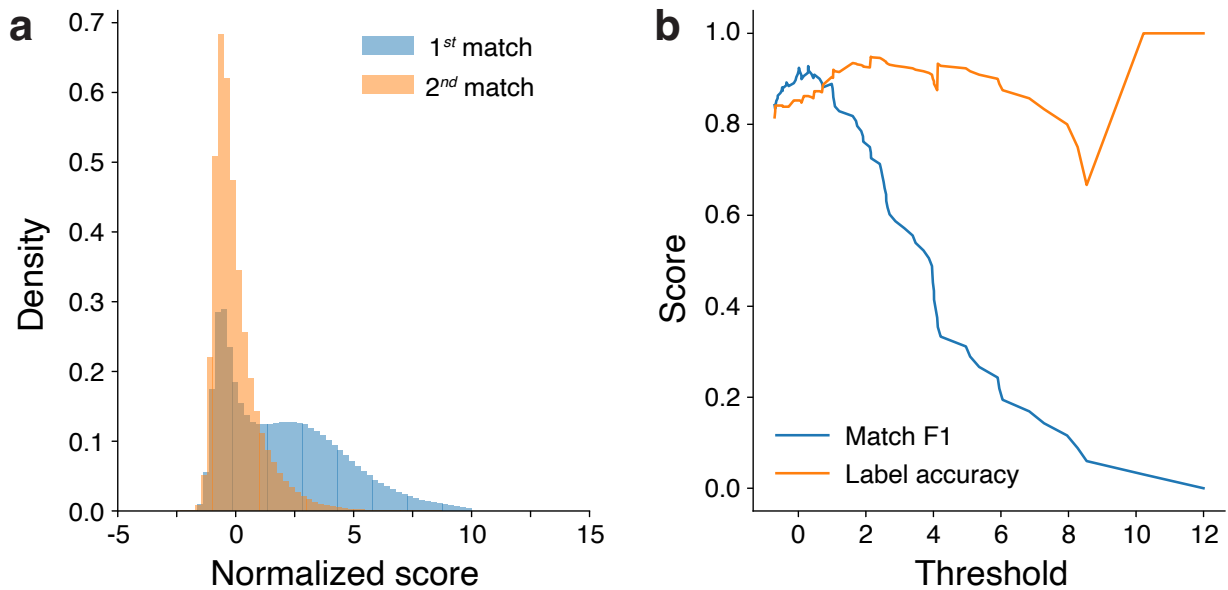

**Figure S2. Matching raw reference string to MAG records.** (**a**) Distribution of normalized score for papers with the first and second highest matching score. The normalization is obtained by calculating the z-score of the raw score for the second-best matched papers for strings of similar word length. The normalized score for second-best matched papers approximately follows a standard Gaussian distribution, yet the first-best matched papers peak near $z = 2$, indicating a large proportion of matchings are significantly more accurate than expected. (**b**) We tuned the threshold and evaluated matching performance on a manually validated subset using two step measures: (1) Whether a string can be matched into a MAG paper (binary classification problem), measured by F1 score, and (2) Conditional on the string successfully matching into MAG both automatically and manually, to what extent are the two matched IDs consistent (label problem), measured by overall accuracy. We find both measures show high predictive power near $z = 2$.

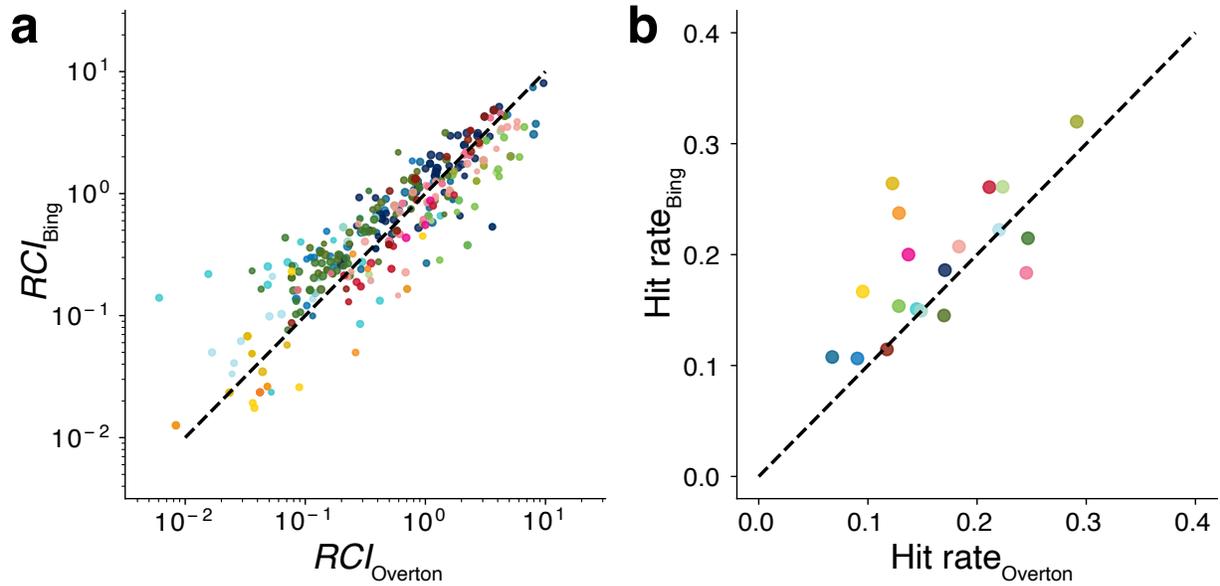

**Figure S3. Comparing Bing data with Overton data.** The x-axis represents values calculated on all policy documents published by U.S.-based sources from the Overton dataset, while the y-axis represents values calculated on all tier-0 .gov domain PDF pages indexed by the Bing search engine. (**a**) Comparing the relative consumption index (*RCI*) for level-1 fields reveals high consistency between the two sources (Spearman correlation $r_s = 0.94$, $p < 0.001$). (**b**) The hit rates of papers used in policy for level-0 fields study are also highly correlated across Overton and Bing (Spearman correlation $r_s = 0.60$, $p = 0.006$). More importantly, the unusually high hit rates are also universal in Overton, where the lowest hit rate across all fields is still 6.7 times the baseline rate of 1%.

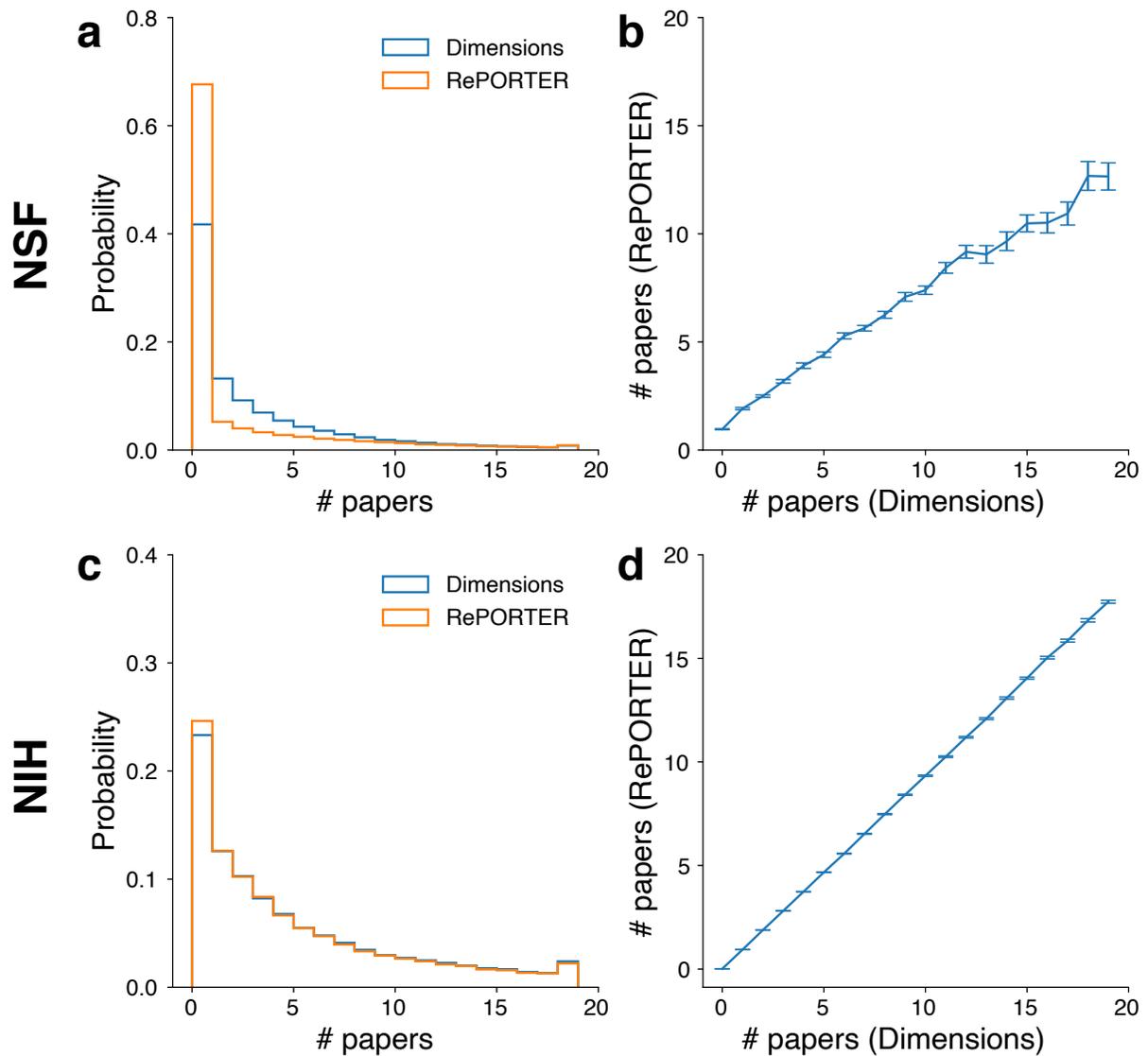

**Figure S4. Comparing papers supported by NSF and NIH grants based on RePORTER and Dimensions data.** (**a**) The distribution of number of papers matched to each NSF grant between 2005 and 2014. (**b**) Number of papers matched to each NSF grant reported by Dimensions and RePORTER data. Dimensions have more papers covered on average. (**c-d**) Same as (**a-b**) but for NIH grants.

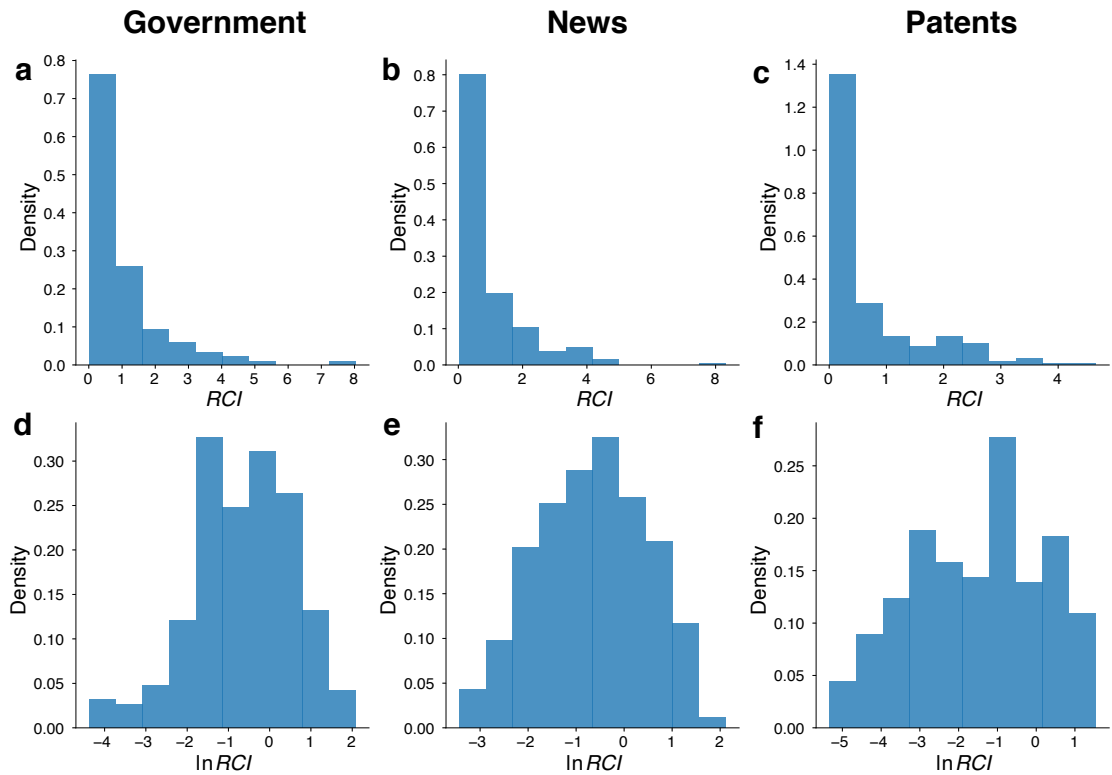

**Figure S5. Distribution of *RCI* index and transformations.** (**a-c**) Histogram of *RCI* for level-1 fields in policy (**a**), news (**b**) and patent (**c**). All three distributions are highly skewed, prompting us to take appropriate transformations before regression analysis. (**d-f**) Histogram of ln *RCI* for level-1 fields in policy (**d**), news (**e**) and patent (**f**). The three distributions are closer to normal distributions after the transformation as compared with (**a-c**).

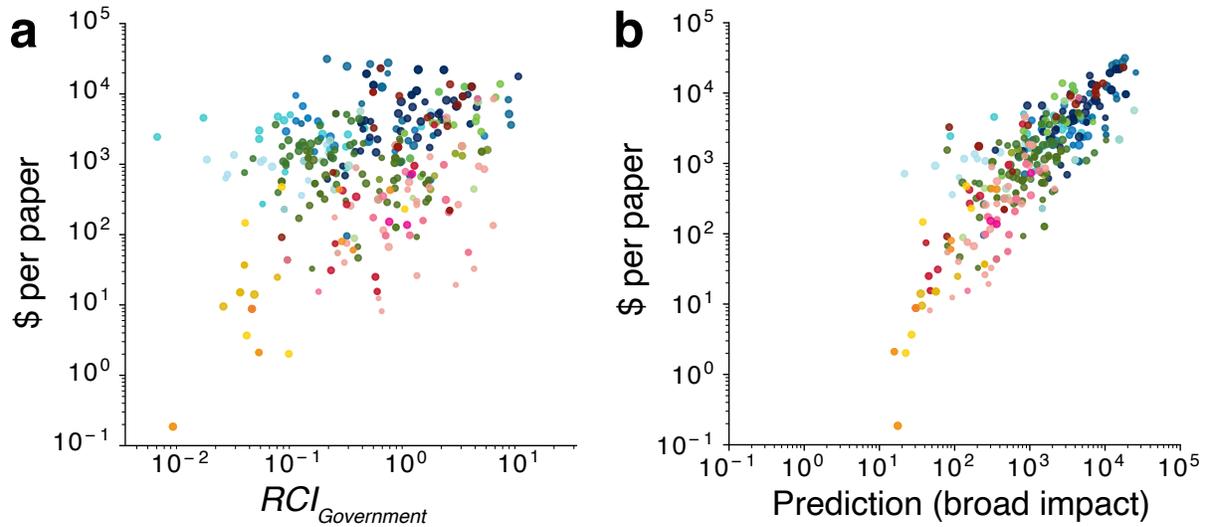

**Figure S6. Robustness checks on policy documents by using Overton data.** (**a**) Average funding per paper across fields is positively correlated with a field's *RCI* index in government (based on Overton data). The relationship remains significant when combined with control variables ($P < 0.001$ after controlling for the number of papers and parent field fixed effects, Table S3). (**b**) Collectively, public uses beyond science strongly predict field level funding per paper.

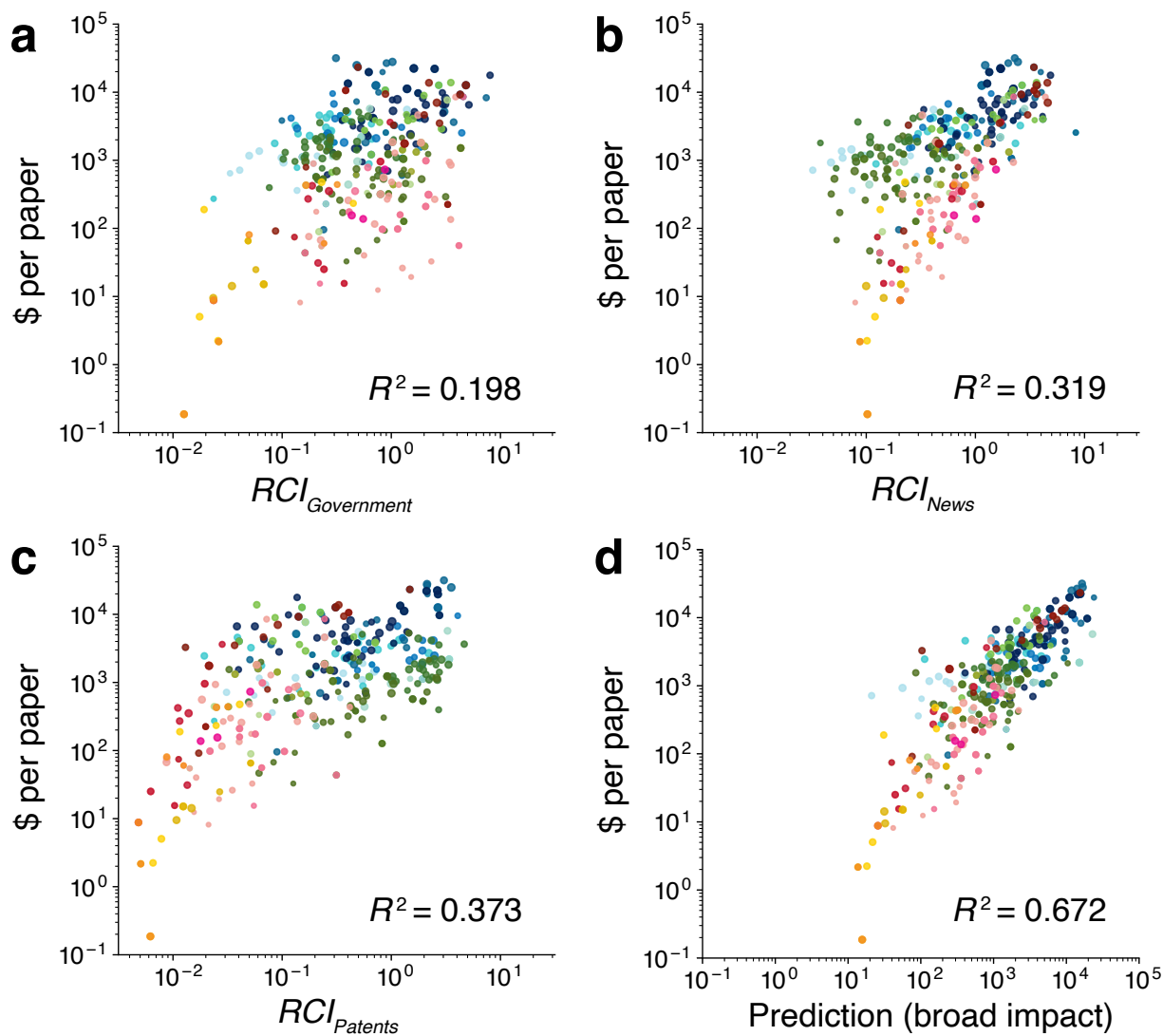

**Figure S7. Public use of science and scientific funding by including all US funding.** (**a-c**) Average funding per paper across fields is positively correlated with a field's *RCI* index in government (**a**), news (**b**) and patenting (**c**). The relationship remains significant when combined with control variables ($P < 0.001$ in all three cases after controlling for the number of papers and parent field fixed effects, Table S5). (**d**) Collectively, public uses beyond science strongly predict field level funding per paper.

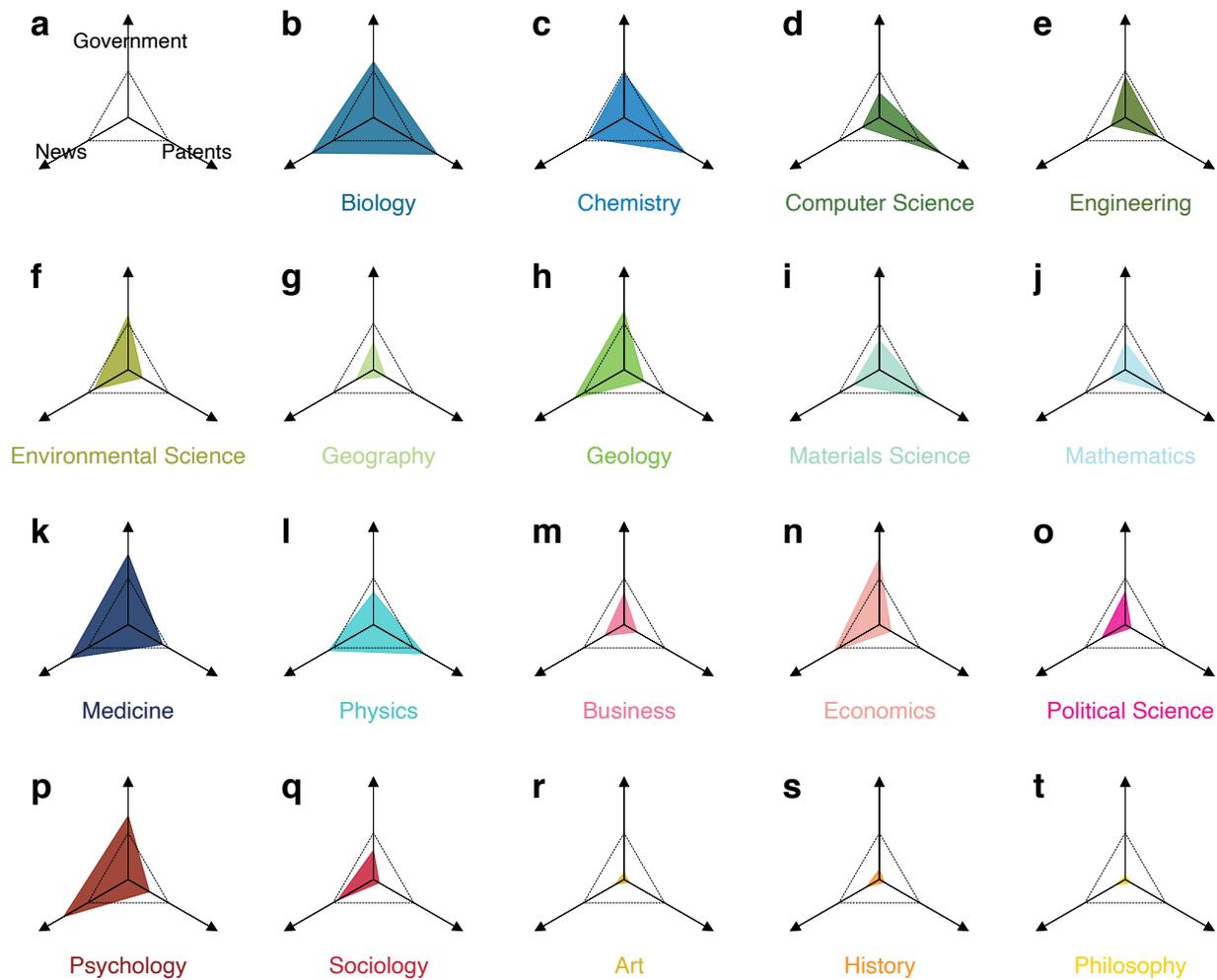

**Figure S8. Robustness checks on definition of academic papers: results using all MAG publications.**

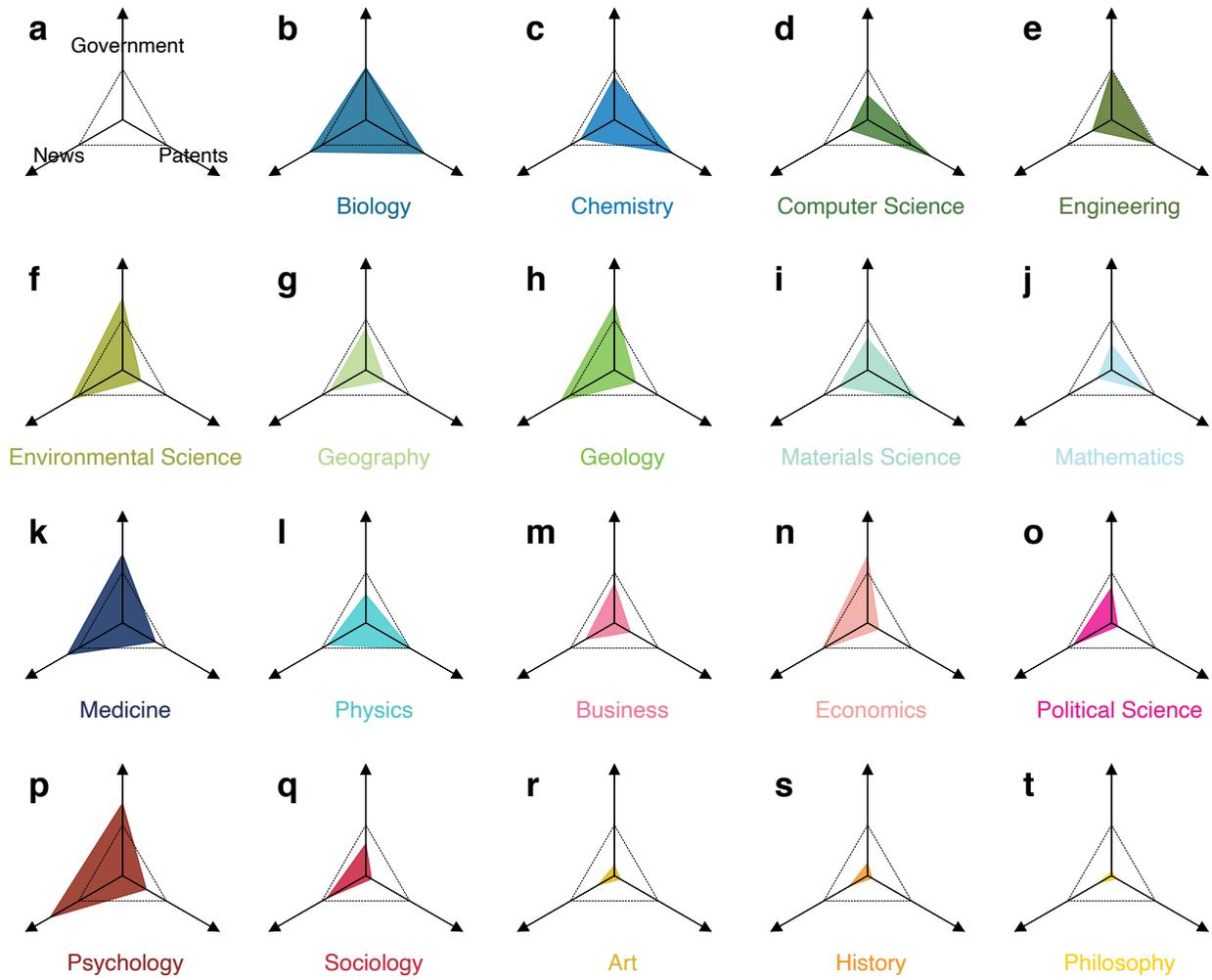

**Figure S9. Robustness checks on definition of academic papers: results only using English publications.**

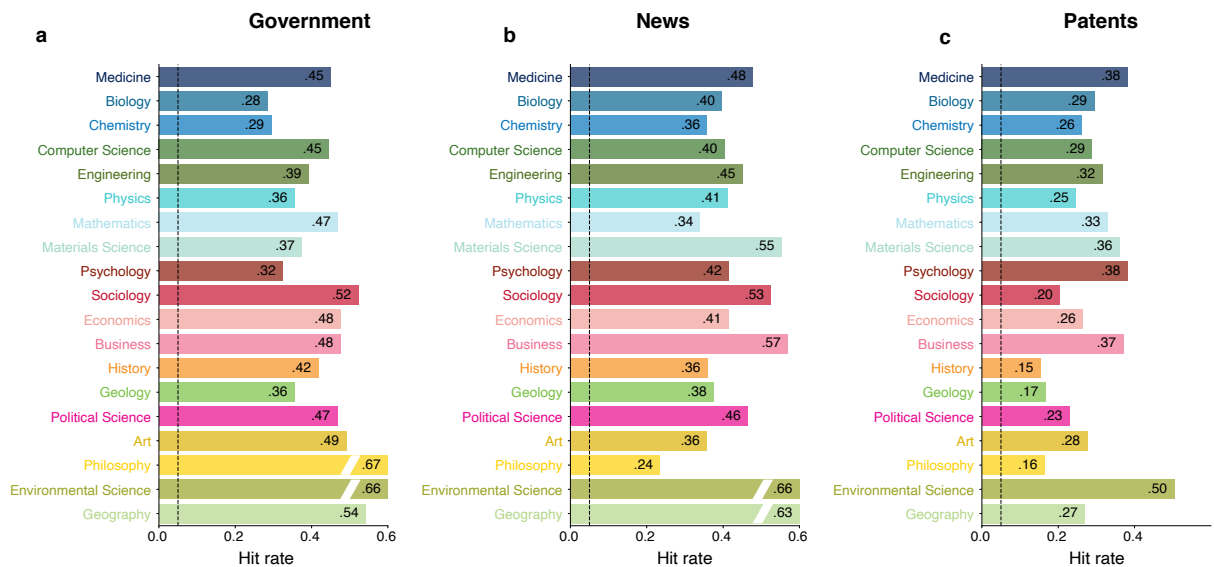

**Figure S10. Robustness checks on paper hit rate.** Paper hit rate for the papers used across 19 fields consumed by government documents (**a**), news media (**b**) and patents (**c**). In all fields, and in all three domains, the consumed papers have hit rates within science many times larger than the baseline rate of 5% (dashed line).

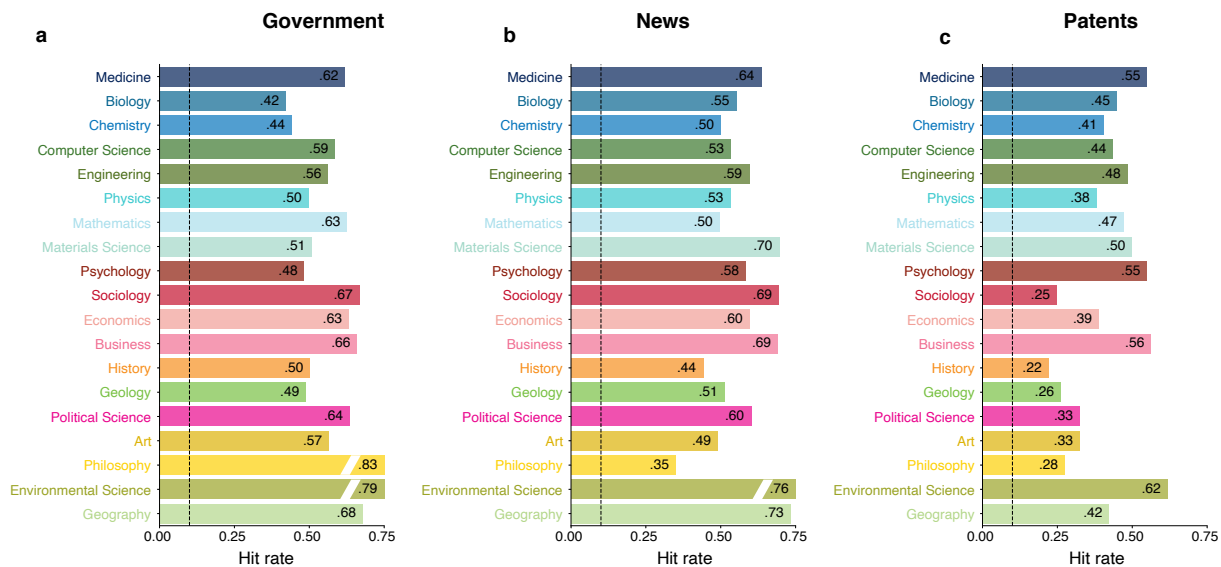

**Figure S11. Robustness checks on paper hit rate.** Paper hit rate for the papers used across 19 fields consumed by government documents (**a**), news media (**b**) and patents (**c**). In all fields, and in all three domains, the consumed papers have hit rates within science many times larger than the baseline rate of 10% (dashed line).

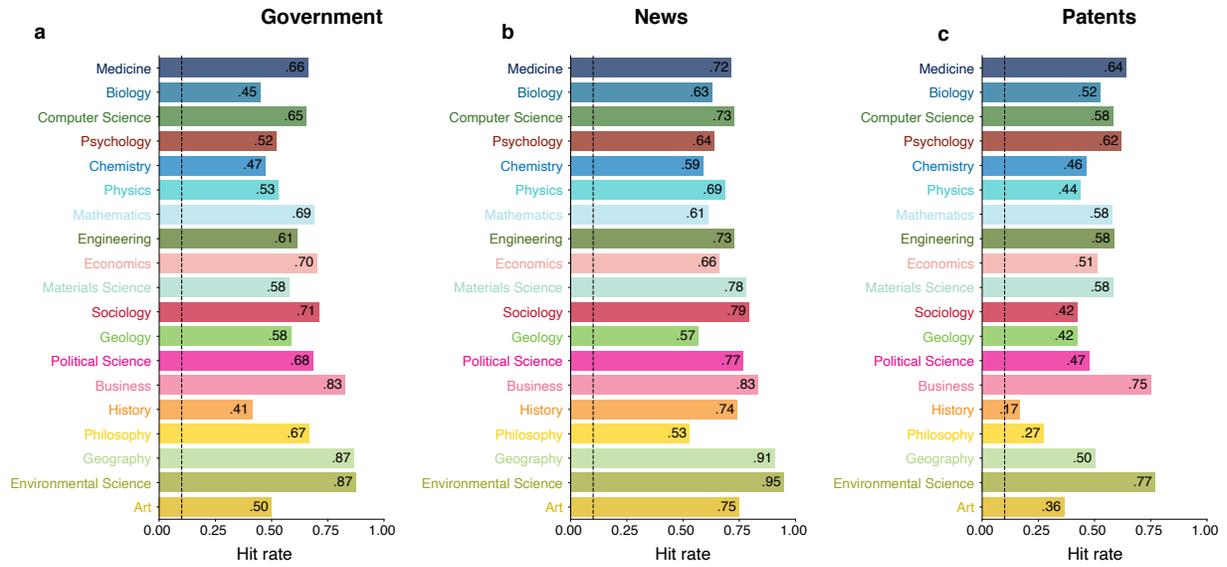

**Figure S12. Robustness checks on paper hit rate for papers produced by U.S.-based researchers.** Paper hit rate for the papers used across 19 fields consumed by government documents (**a**), news media (**b**) and patents (**c**). In all fields, and in all three domains, the consumed papers have hit rates within science many times larger than the baseline rate of 10% (dashed line).

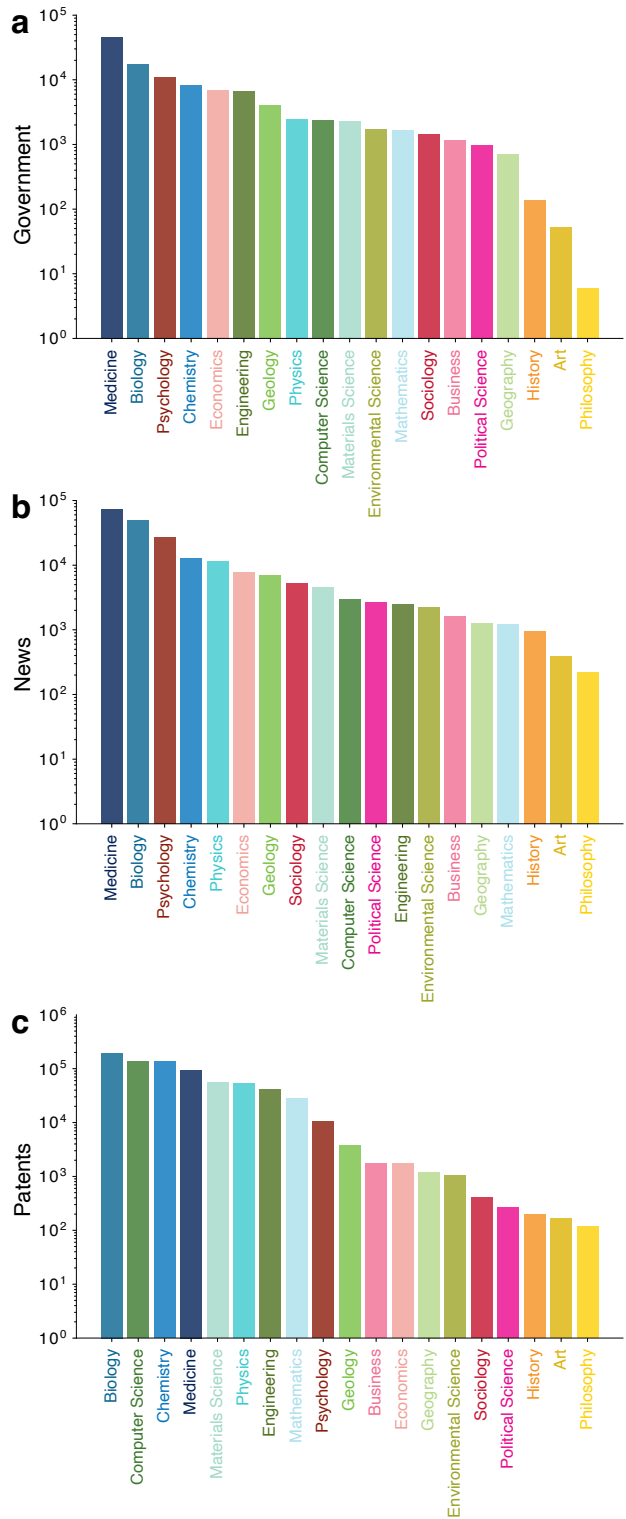

**Figure S13. Ranking of fields by their volume of use.** We rank 19 level-0 fields by number of papers used in government documents (**a**), news media (**b**) and patents (**c**).

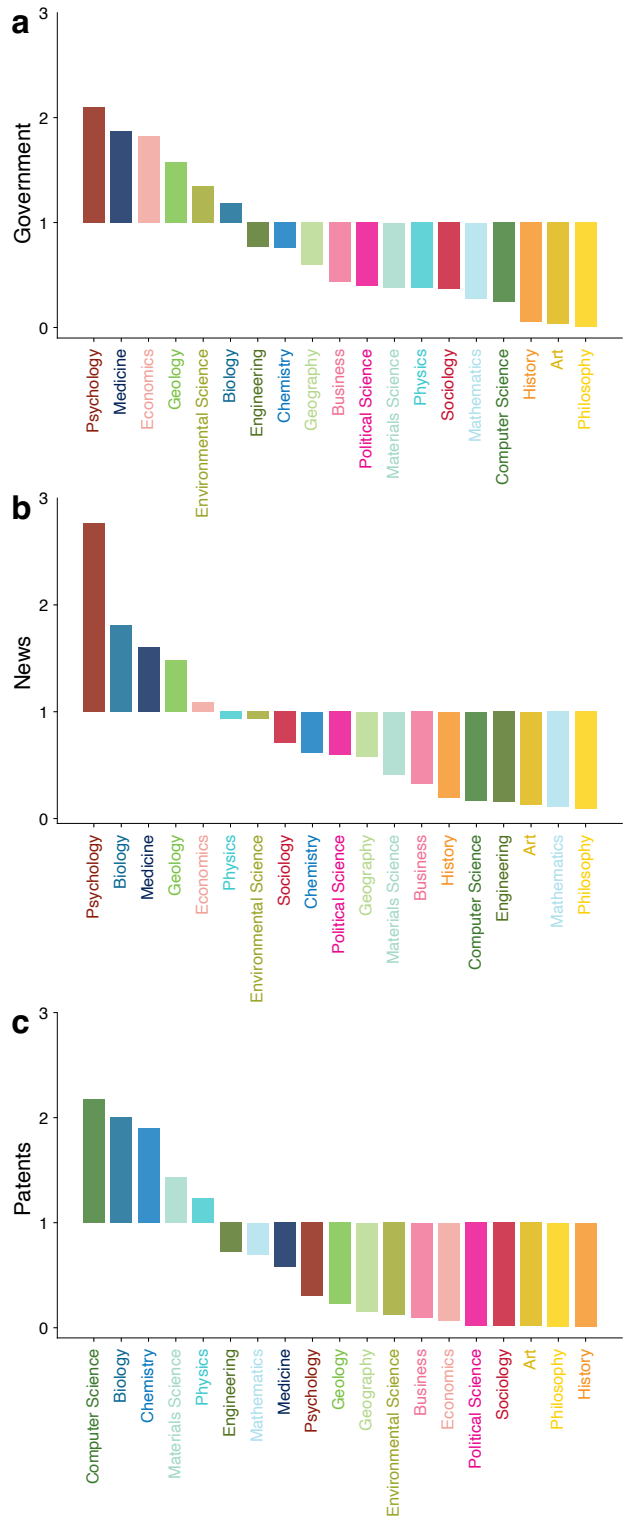

**Figure S14. Ranking of fields by their relative use.** We rank 19 level-0 fields by their relative use (*RCI*) in government documents (**a**), news media (**b**) and patents (**c**).

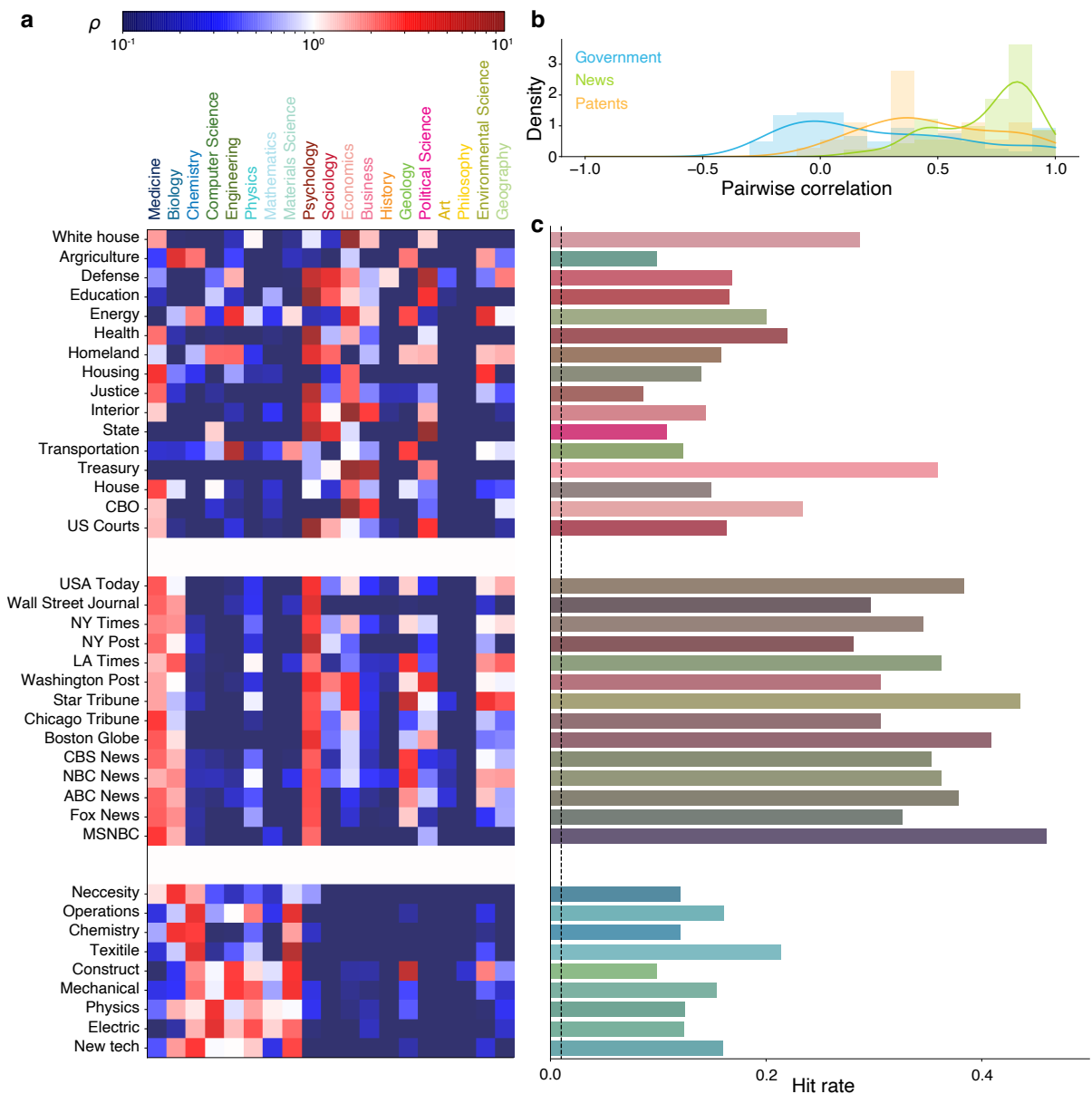

**Figure S15. Use of science across subdomains of government, news, and patents. (a)** Patterns of consumption across subdomains. Subdomains are major departments and entities within the U.S. federal government, major news outlets for U.S. media, and top-level CPC (Cooperative patent classification) technology classes for patents. **(b)** The heterogeneity of scientific consumption across the subdomains in **(a)**. **(c)** For every subdomain, paper hit rates are universally higher than the baseline (dashed line).

30